\documentclass{iopart}
\usepackage{graphicx}
\usepackage{psfrag}
\usepackage{amssymb}
\usepackage{hyperref}
\usepackage[square,comma,sort&compress]{natbib}
\newlength{\sepmod}
\newcommand{\ds}{\displaystyle}

\def\const{\mathrm{const}}
\def\l{\langle}

\def\r{\rangle}

\def\sinh{\mathrm{sinh}}
\def\cosh{\mathrm{cosh}}

\def\arcosh{\mathrm{acosh}}

\def\arccos{\mathrm{arccos}}
\def\d{\mathrm{d}}
\def\RIGHT{\put(  0,0  ){\vector( 1, 0){70}}\put(70,0 ){\line(1,0){30}}}
\def\UP{\put(  0,0  ){\vector( 0, 1){70}}\put(0 ,70){\line(0,1){30}}}
\def\LEFT{\put(100,0  ){\vector(-1, 0){70}}\put(0 ,0 ){\line(1,0){30}}}
\def\DOWN{\put(  0,100){\vector( 0,-1){70}}\put(0 ,0 ){\line(0,1){30}}}
\newcommand{\VERTEX}[5]{\begin{picture}(0,0)
    \put(-100,0   ){#1}       
    \put(   0,0   ){#2}       
    \put(   0,-100){#3}       
    \put(   0,0   ){#4}       
    \put(   0,-180){#5}       
  \end{picture}}

\begin{document}

\title[The square-lattice $F$ model revisited]
{The square-lattice $F$ model revisited: a loop-cluster update scaling study}
\author{M Weigel\dag\ and W Janke\ddag}
\address{\dag Department of Physics, University of Waterloo, 200 University Av W,
  Waterloo, Ontario, N2L~3G1, Canada}
\address{\ddag Institut f\"ur Theoretische Physik, Universit\"at Leipzig, Augustusplatz
  10/11, 04109 Leipzig, Germany}
\eads{\mailto{weigel@boromir.uwaterloo.ca}, \mailto{janke@itp.uni-leipzig.de}}

\begin{abstract}
  The six-vertex $F$ model on the square lattice constitutes the unique example of an
  exactly solved model exhibiting an infinite-order phase transition of the
  Kosterlitz-Thouless type. As one of the few non-trivial exactly solved models, it
  provides a welcome gauge for new numerical simulation methods and scaling
  techniques. In view of the notorious problems of clearly resolving the
  Kosterlitz-Thouless scenario in the two-dimensional {\em XY\/} model numerically,
  the $F$ model in particular constitutes an instructive reference case for the
  simulational description of this type of phase transition. We present a
  loop-cluster update Monte Carlo study of the square-lattice $F$ model, with a focus
  on the properties not exactly known such as the polarizability or the scaling
  dimension in the critical phase. For the analysis of the simulation data,
  finite-size scaling is explicitly derived from the exact solution and plausible
  assumptions. Guided by the available exact results, the careful inclusion of
  correction terms in the scaling formulae allows for a reliable determination of the
  asymptotic behaviour.
\end{abstract}

\submitto{\JPA}

\section{Introduction}
\label{sec:intro}

\begin{figure}[tb]
  \begin{center}
    \setlength{\unitlength}{.00055\textwidth}
    \begin{picture}(1700,350)
      \thicklines
      \put( 100,200){\VERTEX{\RIGHT}{\RIGHT}{\UP}  {\UP}  {1}}
      \put( 400,200){\VERTEX{\LEFT} {\LEFT} {\DOWN}{\DOWN}{2}}
      \put( 700,200){\VERTEX{\RIGHT}{\RIGHT}{\DOWN}{\DOWN}{3}}
      \put(1000,200){\VERTEX{\LEFT} {\LEFT} {\UP}  {\UP}  {4}}
      \put(1300,200){\VERTEX{\RIGHT}{\LEFT} {\DOWN}{\UP}  {5}}
      \put(1600,200){\VERTEX{\LEFT} {\RIGHT}{\UP}  {\DOWN}{6}}
    \end{picture}
    \caption
    {Allowed arrow configurations for the six-vertex model on the square lattice,
      restricted by the {\em ice rule\/}.}
    \label{vertex_conf_fig}
  \end{center}
\end{figure}
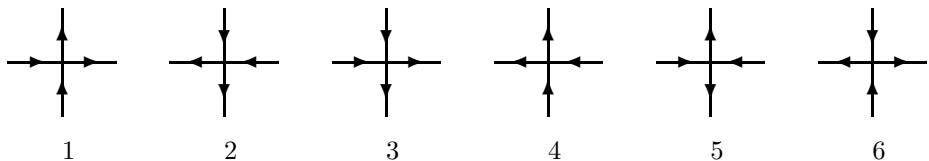

\begin{figure}[tb]
  \centering
  \includegraphics[clip=true,keepaspectratio=true,width=5cm]{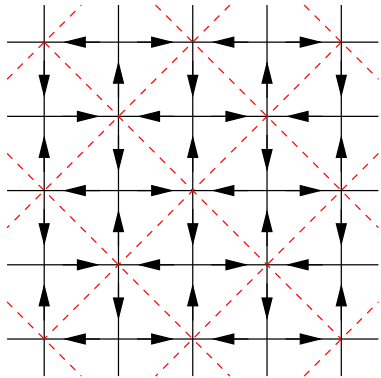}
  \caption
  {Cutout of one of the two anti-ferroelectrically ordered ground states of the
    square-lattice $F$ model. The state consists of vertices $5$ and $6$ at equal
    proportions. The dashed lines indicate one of two tilted sub-lattices of {\em
      ferroelectrical\/} order.}
  \label{sixvertex_groundstates_fig}
\end{figure}

An {\em ice-type\/} or {\em vertex\/} model was first proposed by Pauling
\cite{pauling:35a} as a model for (type I) water ice. It was known that ice forms a
hydrogen-bonded crystal, i.e., the oxygen atoms are located on a four-valent lattice
and the bonding is mediated by one hydrogen atom per bond. Pauling proposed that
there be some non-periodicity in the arrangement of the hydrogen bonds in that the
hydrogen atoms could be located nearer to one or the other end of the bond. This
positioning should satisfy the {\em ice rule\/}, stating that always two of the bonds
are in the ``close'' position and two are in the ``remote'' position with respect to
the considered oxygen atom. Denoting the position of the hydrogen atom by a
decoration of the bond with an {\em arrow\/} pointing to the closer oxygen, this
leads to the arrow configurations depicted in figure \ref{vertex_conf_fig} when for
simplicity placing the oxygens on a square lattice instead of the physically realized
diamond lattice. Generalizing the resulting {\em six-vertex model\/} for square ice,
one assigns energies $\epsilon_i$, $i=1,\ldots,6$, to the vertex configurations
depicted in figure \ref{vertex_conf_fig}, resulting in Boltzmann factors
$\omega_i=\exp(-\beta\epsilon_i)$, where $\beta=1/k_BT$ is the inverse temperature or
coupling. Assuming an overall arrow reversal symmetry (corresponding to the absence
of an external electric field), one abbreviates $a=\omega_1=\omega_2$,
$b=\omega_3=\omega_4$ and $c=\omega_5=\omega_6$. Then, the original ice model
corresponds to the choice $\epsilon_i=0$, $i=1,\ldots,6$, whereas another especially
symmetric version assumes
\begin{equation}
  \label{eq:fmodelweights}
  \epsilon_a=\epsilon_b=1,\;\;\epsilon_c=0\;\;\;\mathrm{resp.}\;\;\;a=b=e^{-\beta},\;\;c=1,
\end{equation}
which is known as the $F$ model of anti-ferroelectrics \cite{rys:63a}, since due to
the choice of weights the vertex configurations 5 and 6 will dominate for low
temperatures, resulting in a ground-state of staggered, anti-ferroelectric order as
depicted in figure \ref{sixvertex_groundstates_fig}.

The six-vertex model as well as the more general eight-vertex models, obtained by
including sink and source vertices with all four arrows pointing in and out,
respectively, have been exactly solved in zero field using transfer matrix
techniques, see reference \cite{baxter:book}. They exhibit rich phase diagrams
featuring first-order and continuous phase transitions as well as multi-critical
points. In particular, the six-vertex $F$ model undergoes an infinite-order
transition of the Berezinskii-Kosterlitz-Thouless (BKT) type to an
anti-ferroelectrically ordered phase and the scaling behaviour of the basic
thermodynamic quantities can be extracted from the closed-form solution. Since there
is no solution of the model in a (staggered) field, however, information about
properties related to the polarization is incomplete.  The same is true for the
correlation function, which can only be evaluated at the so-called free-fermion point
of the model \cite{baxter:70a,baxter:book} (the correlation length, however, is
exactly known for all temperatures, see below). Also, since the solution was obtained
in the thermodynamic limit, information about finite-size scaling (FSS) is not exact,
but must be deduced from scaling arguments. Apart from its prominent position as a
non-trivial solvable model of statistical mechanics, the $F$ model has enjoyed
sustained interest due to its equivalence to the BCSOS surface model
\cite{beijeren:77a}, and hence several {\em dynamical\/} generalizations of the
six-vertex model have been considered \cite{levi:97a}. A six-vertex model with
so-called domain-wall boundary conditions has recently attracted considerable
interest and found numerous applications in counting problems, the quantum inverse
scattering method etc.\ \cite{bogoliubov:02a}.

The Berezinskii-Kosterlitz-Thouless \cite{bkt} scenario of an infinite-order phase
transition induced by the unbinding of vortex pairs in the two-dimensional {\em XY\/}
model has been found exceptionally hard to confirm numerically
\cite{guptaetal,wj:93a,kenna:97a,wj:97b}. This is partially due to the nature of the
infinite-order phase transition itself, which is not easy to distinguish from a
finite-order transition numerically, and the presence of a critical phase, which
render many of the standard FSS techniques less useful. The main trouble, however, is
caused by the presence of logarithmic corrections, expected to be present on general
grounds for a theory with central charge $c=1$ \cite{henkel:book} and explicitly
found from the BKT theory of the model \cite{amitkadanoff}. While a numerical
confirmation of the leading scaling behaviour of the BKT transition has been achieved
in the past decade or so, the resolution of the logarithmic scaling corrections is
still at the forefront of problems amenable to numerical investigation today
\cite{wj:97a,balog:01a,hasenbusch:05a}.

From duality arguments and mapping to Coulomb gas systems, the $F$ model is known to
be asymptotically equivalent to the two-dimensional {\em XY\/} model at criticality. Thus,
apart from being an interesting subject in its own right, a detailed analysis of the
thermal and FSS properties of the six-vertex $F$ model in the critical phase and at
its BKT point serves as a guideline for simulations of the {\em XY\/} model case. Guidance
is been given here through the fact that the exact solution of the $F$ model yields
the leading singularities {\em including the correction terms\/} explicitly, and,
most notably for numerical purposes, the exact critical coupling of the model.
Uncertainties occurring in analyses of the {\em XY\/} model such as systematic errors in
the determination of the transition point or the effect of neglected higher-order
correction terms can be studied rather explicitly for the $F$ model.  Finally, when
it is found here that one has to consider large system sizes and proceed carefully
when including correction terms into the fits, this situation should also be put into
relation with the case of an $F$ model placed on an annealed ensemble of {\em random
  lattices\/} considered recently \cite{weigel:04b}.  Guided by the present
investigation, this case has to be analyzed even more carefully due to an additional
fractality of the lattices, which reduces the effective linear extent of the amenable
lattice sizes, thus increasing finite-size effects even further.

The other paradigm example of an exactly solved non-trivial model of statistical
mechanics, the two-dimensional Ising model, has served as a benchmark system and
playground for new ideas in the theory of critical phenomena as well as for new
algorithms in computer simulations in an overwhelming number of studies, and almost
all of its aspects have been investigated (but not necessarily understood). In
contrast, for the case of vertex models only rather recently efficient cluster-update
Monte Carlo algorithms have been developed
\cite{evertz:loop,barkema:98a,syljuasen:04a}, mainly with the mapping of vertex
models on quantum chains in mind, and some simulations of special aspects of the
six-vertex model, such as dynamical critical exponents of the considered algorithms
\cite{evertz:93b,barkema:98a}, properties of the equivalent surface models
\cite{mazzeo}, matching of renormalization-group flows with the {\em XY\/} model
\cite{hasenbuschpinn}, or the case of domain-wall boundary conditions
\cite{syljuasen:04a} have been analyzed. A systematic thermal and FSS study of the
$F$ model in the critical high-temperature phase, at the critical point and its
low-temperature vicinity including the analysis of the logarithmic correction terms,
however, is to our best knowledge lacking so far.

The rest of the paper is organized as follows. In section 2 we outline the extent of
exact knowledge about the phase diagram and the occurring transitions of the
six-vertex model and the $F$ model in particular and give an overview over scaling at
a BKT point in general. After a short description of the simulational setup used,
section 3 contains a report of the analysis of the simulation data, comprising the
FSS analyses of the critical-point thermodynamic properties (where the corresponding
FSS relations are explicitly derived from the closed-form solution), an investigation
of the behaviour in the critical phase as well as a thermal scaling analysis in the
low-temperature phase of the model. Finally, section 4 contains our conclusions.

\section{Analytical Results}

\subsection{Exact solution and phase diagram}

\begin{figure}[tb]
  \centering
  \includegraphics[clip=true,keepaspectratio=true,width=5.5cm]{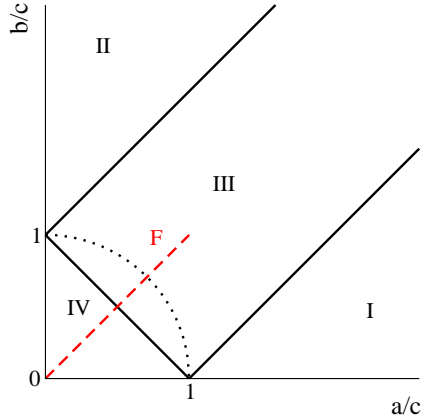}
  \caption
  {The phase diagram of the square-lattice, zero-field six-vertex model in terms of
    the re-scaled weights $a/c$ and $b/c$. Phase boundaries are indicated by solid
    lines. The dashed line denotes the parameter range of the $F$ model. The dotted
    line corresponds to the free-fermion line $\Delta=0$.}
  \label{sixvertex_phase_diagram}
\end{figure}

The square-lattice, zero-field six-vertex model has been solved exactly in the
thermodynamic limit by means of the Bethe ansatz by Lieb \cite{lieb} and Sutherland
\cite{sutherland:67a}. The analytic structure of the free energy is most conveniently
parametrized in terms of the reduced coupling
\begin{equation}
  \Delta=\frac{a^2+b^2-c^2}{2ab},
  \label{delta_def}
\end{equation}
such that the free energy takes a different analytic form depending on whether
$\Delta<-1$, $-1<\Delta<1$ or $\Delta>1$. This leads to a phase diagram of the model
consisting of four distinct phases as shown in figure \ref{sixvertex_phase_diagram}.
The phases I and II are both characterized by $\Delta>1$, thus corresponding to the
same analytic form of the free energy; they represent ferroelectrically ordered
phases, the ground-states being related to each other through a global rotation by
$\pi/2$. In these phases, the system exhibits the peculiarity of sticking to the
respective ground-states also for non-zero temperatures. The intermediate case
$-1<\Delta<1$, corresponding to phase III, includes the infinite temperature point
$a=b=c=1$ and thus belongs to a disordered phase, which turns out to be massless,
i.e., it exhibits algebraic correlations throughout. This latter effect can be traced
back to the fact that the six-vertex model corresponds to a critical surface in the
phase diagram of the eight-vertex model. Finally, for $\Delta<-1$ one has $c>a+b$,
resulting in the anti-ferroelectric order of phase IV described above for the $F$
model.  The parameter space of the $F$ model is restricted to the dashed line
connecting phases III and IV indicated in figure \ref{sixvertex_phase_diagram}. The
dotted line of figure \ref{sixvertex_phase_diagram} indicates the curve $\Delta=0$,
where the six-vertex model is equivalent to a system of free fermions and an exact
solution is even possible in the presence of a staggered field
\cite{baxter:70a,baxter:book}.  The nature of the transitions between the phases
I--IV can be extracted from the exact solution
\cite{lieb,sutherland:67a,baxter:book}. Crossing the phase boundaries I $\rightarrow$
III and II $\rightarrow$ III one finds discontinuities corresponding to first-order
transitions. The transition III $\rightarrow$ IV, on the other hand, is peculiar in
that all the temperature derivatives of the free energy exist and vanish
exponentially as the transition is approached. These are the properties of a BKT
phase transition to be detailed below in section \ref{sec:kt}.

While the ferroelectrically ordered phases I and II exhibit a {\em plain\/}
polarization, which can be used as an order parameter for the corresponding
transition, the anti-ferro\-electric order of phase IV is accompanied by a {\em
  staggered\/} polarization with respect to a sub-lattice decomposition of the square
lattice. This is equivalent to a mutually inverse plain polarization on two tilted,
square sub-lattices as indicated in figure \ref{sixvertex_groundstates_fig}.  An
order parameter for the corresponding transition can be defined by introducing
overlap variables $\sigma_i$ for each vertex $i$ such that \cite{baxter:book},
\begin{equation}
  \sigma_i = v_i\ast v_i^0,
  \label{staggered_polarization}
\end{equation}
where $v_i^0$ denotes the anti-ferroelectric ground-state configuration depicted in
figure \ref{sixvertex_groundstates_fig} and the product ``$\ast$'' denotes the
overlap given by
\begin{equation}
  v\ast v' \equiv \sum_{k=1}^4 A_k(v) A_k(v'),
\end{equation}
where $k$ numbers the four edges around each vertex and $A_k(v)$ should be $+1$ or
$-1$ depending on whether the corresponding arrow of $v$ points out of the vertex or
into it. The thus defined {\em spontaneous staggered polarization\/} $P_0\equiv
\l\sigma_i\r/2=\l\sigma\r/2$ constitutes an order parameter for the antiferroelectric
transition III $\rightarrow$ IV.

As indicated above, the $F$ model orders anti-ferroelectrically at $\Delta=-1$,
corresponding to a critical coupling $\beta_c=\ln 2$. From the exact solution
\cite{lieb,sutherland:67a,baxter:book}, the model's asymptotic free energy per site
in the low-temperature phase $\beta>\beta_c$ can be written as
\begin{equation}
  \label{eq:f_low}
  f^\mathrm{low}(\lambda) = \beta-\frac{\lambda}{2}-\sum_{m=1}^\infty
  \frac{\exp(-m\lambda)\,\sinh(m\lambda)}{m\,\cosh(m\lambda)},
\end{equation}
where $\lambda=\arcosh[\frac{1}{2}\exp(2\beta)-1]$. On the high-temperature side
$\beta<\beta_c$ it takes a different analytic form and has the following integral
representation,
\begin{equation}
  \label{eq:f_high}
  f^\mathrm{high}(\mu) = \beta-\frac{1}{4\mu}\int_0^\infty\frac{\d t}{\cosh(\pi t/2\mu)}
  \ln\left(\frac{\cosh\,t-\cos 2\mu}{\cosh\,t-1}\right),
\end{equation}
where $\mu=\arccos[\frac{1}{2}\exp(2\beta)-1]$. The correlation length is given by
the two equivalent expressions
\begin{equation}
  \label{eq:corrlength}
  \exp[-1/\xi(\lambda)] = 2 x^{1/4}\prod_{m=1}^\infty\left(
      \frac{1+x^{2m}}{1+x^{2m-1}}\right)^2 = 
    \prod_{m=1}^\infty\left(\frac{1-y^{2m-1}}{1+y^{2m-1}}\right)^2,
\end{equation}
where $x=\exp(-2\lambda)$ and $y=\exp(-\pi^2/2\lambda)$ are ``dual'', conjugate nomes
of an elliptic function, yielding two different representations being rapidly
convergent for large $\lambda$ (first form) and small $\lambda$ (second form),
respectively.  Although the general $F$ model has not been solved in a staggered
electric field, as a single result the spontaneous staggered polarization is known
exactly for all inverse temperatures $\beta > \beta_c$ \cite{baxter:73b},
\begin{equation}
  \label{eq:polarization}
  \fl
  P_0(\lambda)^{1/2} = \prod_{n=1}^\infty \tanh(n\lambda) = 1+2\sum_{n=1}^\infty(-1)^nx^{n^2} =
  \left(\frac{2\pi}{\lambda}\right)^{1/2}\sum_{n=1}^\infty y^{(n-\frac{1}{2})^2},
\end{equation}
where again the first two forms are rapidly convergent for large $\lambda$, away from
criticality, and the third form converges fast close to the critical point
$\lambda=0$. As a proper order parameter, the spontaneous polarization $P_0$ vanishes
identically in the critical high-temperature phase $\beta < \beta_c$.

\subsection{The Berezinskii-Kosterlitz-Thouless phase transition}
\label{sec:kt}

As stated, the $F$ model undergoes an finite-order phase transition of the BKT type
at $\beta_c=\ln 2$. For later reference, let us shortly bring to mind the basic
features of the BKT scenario for the two-dimensional {\em XY\/} model \cite{bkt},
which forms the paradigmatic case of an infinite-order phase transition, albeit the
exact solution of the $F$ model was published a couple of years earlier. As a
consequence of the Mermin-Wagner-Hohenberg theorem \cite{merminwagner}, the
two-dimensional {\em XY\/} model cannot develop an ordered phase with a non-vanishing
value of a locally defined order parameter for non-zero temperatures\footnote{Note,
  however, that on a {\em finite\/} lattice, the magnetization attains a non-zero
  value in the low-temperature phase, cf.~reference~\cite{bramwell}.}. Nevertheless,
it undergoes a finite-temperature phase transition resulting from the unbinding of
{\em vortex pairs\/} superimposed on an effective spin-wave behaviour of the
low-temperature phase. Above the critical temperature, spin-spin correlations decay
exponentially,
\begin{equation}
  G(r) \sim e^{-r/\xi(T)},\;\;\; T\downarrow T_c,
\end{equation}
while below $T_c$ long-range correlations are encountered,
\begin{equation}
  G(r) \sim r^{-\eta(T)},\;\;\;T\le T_c,
\end{equation}
such that the correlation length $\xi(T)=\infty$ for all $T\le T_c$ and the massless
low-temperature phase corresponds to a critical line terminating in the critical
point $T_c$ \cite{bkt}. The critical exponent $\eta=\eta(T)$ varies continuously
along this critical line, with $\eta_c=\eta(T_c)=1/4$. Approaching the critical point
$T_c$ from above, the correlation length diverges {\em exponentially\/} instead of
{\em algebraically\/} as for a usual continuous phase transition,
\begin{equation}
  \xi(T) \sim \exp(a/t^{\rho}),\;\;\;t>0,
  \label{exponential_xi_sing}
\end{equation}
where $t=(T-T_c)/T_c$ and $\rho=1/2$. The behaviour of further observables at the
transition point can be conveniently expressed in terms of this singularity of the
correlation length.  In particular, the magnetic susceptibility diverges as
\begin{equation}
  \chi(T)\sim\xi^{\gamma/\nu}=\xi^{2-\eta_c},\;\;\;T\downarrow T_c.
\end{equation}
The specific heat, on the other hand, is only very weakly singular, behaving as
(omitting a regular background contribution)
\begin{equation}
  C_v \sim \xi^{-2}.
  \label{KT_cv_sing}
\end{equation}
Finite-size scaling analyses of the BKT transition are hampered by the occurring
essential singularities and the presence of a critical phase. As a consequence of the
latter, magnetic observables such as the susceptibility do not exhibit maxima in the
vicinity of the critical point, which otherwise could be used for an estimation of
the transition temperature from finite systems. For the same reason, also the Binder
parameter requires a more careful treatment than at a standard second-order phase
transition \cite{bramwell,loison:99a}. Nevertheless, the general arguments for
finite-size shifting and rounding remain valid, such that suitably defined
pseudo-critical points $T^{\ast}(L)$ for systems with linear extent $L$ scale to the
critical point $T_c$ as \cite{barber:domb}
\begin{equation}
  [T^{\ast}(L)-T_c]/T_c\sim(\ln L)^{-1/\rho},
  \label{KT_FSS_shifts}
\end{equation}
cf.\ equation (\ref{exponential_xi_sing}), since sufficiently close to the critical
point the growth of the correlation length becomes limited by the linear extent $L$
of the system. Correspondingly, $\xi$ can be replaced by $L$ (neglecting corrections
to scaling for the time being) to yield the FSS law
\begin{equation}
  \chi(T_c,L) \sim L^{\gamma/\nu}=L^{2-\eta_c},
\end{equation}
which for $\eta_c=1/4$ predicts a rather strong divergence. On finite lattices, the
specific heat is found to exhibit a smooth peak, which is however considerably
shifted away from the critical point into the high-temperature phase and does not
scale as the lattice size is increased \cite{barber:domb}. Thus, with the main
strengths of FSS being not exploitable for the BKT phase transition, the focus of
numerical analyses of the {\em XY\/} and related models has been on {\em thermal\/}
scaling, see, e.g., references \cite{guptaetal,wj:93a}. In addition, renormalization
group analyses predict {\em logarithmic corrections\/} to the leading scaling
behaviour \cite{amitkadanoff}, as expected for a theory of central charge $c=1$,
which have been found exceptionally hard to reproduce numerically due to the presence
of higher order corrections of comparable magnitude (for the accessible lattice
sizes) \cite{kenna:97a,wj:97b,hasenbusch:05a}.

From the exact solution of the square-lattice $F$ model, equations
(\ref{eq:f_low})--(\ref{eq:polarization}), one extracts the asymptotic behaviour in
the vicinity of the critical point $\beta_c=\ln 2$. Approaching the critical point
from the low-temperature side, $\lambda\downarrow 0$, the singular part of the free
energy density (\ref{eq:f_low}) and the correlation length (\ref{eq:corrlength})
behave as
\begin{equation}
  \begin{array}{rcl}
    f_\mathrm{sing}(\lambda) & \sim & 4k_BT_c\exp(-\pi^2/\lambda), \\
    \xi^{-1}(\lambda) & \sim & 4\exp(-\pi^2/2\lambda).
  \end{array}
  \label{fmodel_xifscaling}
\end{equation}
Since $\lambda$ goes as $\lambda\sim (-t)^{1/2}$ for $t\uparrow0$, this exactly
corresponds to the essential singularity described above for the BKT transition of
the two-dimensional {\em XY\/} model with $\rho=1/2$. The specific heat has the
weakly singular contribution $C_v\sim \xi^{-2}$ as expected. Concerning properties
related to the order parameter, the situation for the $F$ model is somewhat different
from that of the {\em XY\/} model. The order parameter (\ref{eq:polarization}) is
non-vanishing for finite temperatures in the ordered phase\footnote{Note that the
  Mermin-Wagner-Hohenberg theorem \cite{merminwagner} does not apply to the $F$ model
  with its discrete symmetry.}. Thus, the corresponding staggered anti-ferroelectric
polarizability $\chi$ shows a clear peak in the vicinity of the critical point for
finite lattices.  However, in the limit $L\rightarrow\infty$ the polarizability
diverges throughout the whole critical high-temperature phase. Note that compared to
the {\em XY\/} model the r\^oles of high- and low-temperature phases are exchanged in
this respect, as expected from duality \cite{savit:80a}. The spontaneous polarization
(\ref{eq:polarization}) scales as
\begin{equation}
  P_0(\lambda) \sim \lambda^{-1}\exp(-\pi^2/4\lambda) \sim \xi^{-1/2}\ln\xi
  \label{fmodel_pscaling}
\end{equation}
as $\lambda\downarrow0$, implying $\beta/\nu=1/2$. Assuming the Widom-Fisher scaling
relation $\alpha+2\beta+\gamma=2$ to be valid\footnote{Although the BKT transition is
  characterized by essential singularities and thus the conventional critical
  exponents are meaningless, one can re-define them by considering scaling as a
  function of the correlation length $\xi$ instead of the reduced temperature $t$
  \cite{suzuki:74a}. The exponents $\alpha$, $\beta$ and $\gamma$ used here and in
  the following should be understood in this sense. The exponent $\rho$, however, has
  its special meaning defined by (\ref{exponential_xi_sing}).}, from equations
(\ref{fmodel_xifscaling}) and (\ref{fmodel_pscaling}) Baxter conjectured the
following scaling of the zero-field staggered polarizability \cite{baxter:73b},
\begin{equation}
  \chi(\lambda) \sim \lambda^{-2}\exp(\pi^2/2\lambda) \sim \xi(\ln\xi)^2,
  \label{polariz_scaling_conj}
\end{equation}
which implies $\gamma/\nu = 2-\eta_c = 1$, obviously different from the {\em XY\/} model
result $\eta_c=1/4$. Since the whole high-temperature phase is critical, scaling of
the polarizability is expected throughout this phase. In fact, for the free-fermion
case $\Delta=0$ or $\beta=\beta_f\equiv (\ln 2)/2$, which is exactly solvable in a
staggered field \cite{baxter:70a}, a logarithmic divergence of the polarizability is
found, implying $2-\eta_f=0$.  More recently, the behaviour of $\eta$ in the critical
phase of the $F$ model has been conjectured from scattering methods to follow the
form \cite{youngblood,bogoliubov:84a,mazzeo}
\begin{equation}
  \eta(\Delta) = \pi/\mathrm{arccos}\Delta.
  \label{eq:delta_conjecture}
\end{equation}
This is in agreement with the exact results for the critical $F$ model at $\Delta=-1$
and the free-fermion case at $\Delta=0$. Additionally, the pure ice model at
$\Delta=1/2$ is known to have a ``dipolar'' correlation function with $\eta = 3$ as
predicted by (\ref{eq:delta_conjecture}) \cite{youngblood,henley:05a}. Note that,
since the dual relation to the {\em XY\/} model is only valid at criticality and the
{\em XY\/} model magnetization is not equivalent to the polarizability of the $F$
model, this result is not simply related to the exponent $\eta$ of the {\em XY\/}
model in its critical low-temperature phase, which actually {\em decreases} as one
moves into the critical phase, see, e.g., references \cite{amitkadanoff,berche:04a}.

The common occurrence of a BKT type phase transition for the {\em XY\/} and $F$ models is no
coincidence. In fact, it can be shown that the critical points of both models are
asymptotically dual to each other \cite{savit:80a}. This can be seen by noting that
the Villain representation of the {\em XY\/} model \cite{villain:75a} is dually equivalent
to a model of the solid-on-solid (SOS) type known as the discrete Gaussian model
\cite{knops:77a}, which in turn, as typical for SOS models, can be mapped onto the
Coulomb gas \cite{nienhuis:domb}. The $F$ model, on the other hand, also has a
height-model representation known as the BCSOS (body-centered SOS) model
\cite{beijeren:77a}, which is itself asymptotically equivalent to the Coulomb gas.
Alternatively, the stated equivalence can be seen from the loop representation of the
O($n$) vector model \cite{domany:81a}, which for the critical O($2$) model yields a
close-packed loop ensemble equivalent to that of the loop representation of the
critical $F$ model \cite{baxter:76a}. The apparent discrepancy regarding the magnetic
exponents $\beta/\nu$ and $\gamma/\nu$ between the {\em XY\/} and $F$ models, on the other
hand, is not an indicator of different universality classes of the models, but
reflects the fact that the $F$ model staggered polarizability is not equivalent to
the magnetic susceptibility of the {\em XY\/} model.

\section{A Loop-Cluster Update Scaling Study}

\subsection{Simulation Setup}

For an analysis of the six-vertex $F$ model via Monte Carlo simulations, a suitable
simulation update scheme has to be devised. Since the focus here lies on the
investigation of the vicinity of the BKT transition and the critical phase of the
model, all local updates will suffer from the severe critical slowing down with
dynamical critical exponent $z\approx 2$ expected at or close to criticality.
Fortunately, a fully-fledged framework of cluster algorithms has been constructed for
the simulation of the six- and eight-vertex models, mainly motivated by their
equivalence with the Trotter-Suzuki decomposition of quantum spin chains.  Here, we
apply the so-called {\em loop-cluster algorithm\/} \cite{evertz:03a}, which operates
on a representation of the vertex model by polygons consisting of the lattice edges
and induced by a stochastic breakup of the lattice vertices, for details see
reference \cite{evertz:03a}.  For the case of the $F$ model at criticality, a
reduction of critical slowing down to $z=0.71(5)$ has been reported
\cite{evertz:93a}.

Simulations were performed for square lattices with periodic boundary conditions,
measurements were taken after each multi-cluster loop-update step due to the small
autocorrelation times observed. To enable a proper FSS analysis, for the
investigation of the BKT point two main series of simulations were performed; one
around the peak locations of the staggered anti-ferroelectric polarizability for
sizes $L=16$, 24, 32, 46, 64, 92, 128, 182, and 256 and another at the asymptotic
critical coupling $\beta_c=\ln 2=0.6931\ldots$ with additional lattice sizes of
$L=364$, 512, 726 and 1024. To examine the behaviour in the critical phase,
additional series of simulations have been performed at fixed temperatures
$\beta<\beta_c$ with lattice sizes identical to those at $\beta_c$. Per simulation,
after equilibration a total of between $1\times 10^5$ and $2\times 10^5$ measurements
were taken.

\subsection{Results of the Finite-Size Scaling Analysis}

\subsubsection{Non-scaling of the specific heat.}

\begin{figure}[tb]
  \centering
  \includegraphics[clip=true,keepaspectratio=true,width=9.5cm]{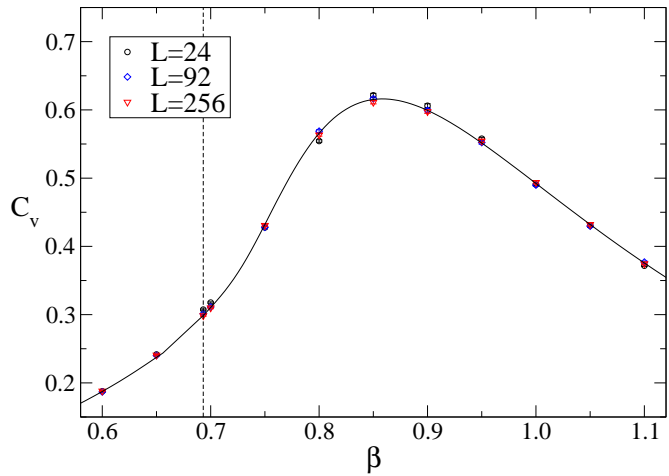}
  \caption
  {Non-scaling of the specific heat $C_v$ of the square-lattice $F$ model. For
    clarity, simulation results from only three lattice sizes are shown. The solid
    line denotes the exact asymptotic result found from the free energy density of
    equations (\ref{eq:f_low}) and (\ref{eq:f_high}). The dashed vertical line marks
    the infinite-volume critical point $\beta_c = \ln 2 = 0.6931\dots$.}
  \label{reg_F_heat}
\end{figure}

The specific heat is defined by
\begin{equation}
  \label{eq:heat_def}
  C_v = \beta^2[\l E^2\r-\l E\r^2]/L^2,
\end{equation}
with the internal energy $E$ of a vertex configuration,
\begin{equation}
  E = \sum_i E(v_i),\;\;\;E(v_i) \in \{\epsilon_1,\ldots,\epsilon_6\},
\end{equation}
where $v_i$ denotes the configuration of vertex $i$ of the lattice. It exhibits a
broad peak shifted away from the critical point into the low-temperature phase
\cite{lieb:domb}\footnote{Recall that the specific heat of the 2D {\em XY\/} model
  exhibits a peak in the {\em high-temperature\/} phase \cite{wj:93a}, as expected
  from duality.}. The essential singularity predicted by equation (\ref{KT_cv_sing})
cannot in general be resolved, since it is covered by the presence of non-singular
background terms. Thus, the {\em non-scaling\/} of a broad specific-heat peak
(together with a scaling of the susceptibility or polarizability to be considered
below) is commonly taken as a first good indicator for a phase transition to be of
the BKT type \cite{barber:domb}.  Indeed, this is what is found from the simulation
data as is shown in figure \ref{reg_F_heat}. No scaling is visible, apart from very
minor deviations for the smallest lattice sizes and close to criticality. All data
points collapse onto a single curve, which is identical to the exact asymptotic
behaviour of $C_v$ extracted from the free energy density of equations
(\ref{eq:f_low}) and (\ref{eq:f_high}) as displayed in figure \ref{reg_F_heat} for
comparison. In particular, at the critical point $\beta_c=\ln 2$ we find for the
internal energy $U=\l E\r$ and the specific heat $C_v$ for the $1024^2$ lattice,
\begin{equation}
  \label{eq:critical_heat}
  \begin{array}{rcl}
    U(\beta_c) & = & 0.333335(4),\\
    C_v(\beta_c) & = & 0.3005(15),
  \end{array}
\end{equation}
in perfect agreement with the exact results $U(\beta_c)=1/3$ and $C_v(\beta_c)=28(\ln
2)^2/45\approx0.2989$ \cite{lieb:domb}.

\subsubsection{The critical coupling.\label{sec:critical_coupling}}

For an independent determination of the critical coupling $\beta_c$ from the
simulation data, we exploit the fact that the maxima of the staggered polarizability
for finite lattices should be shifted away from the critical point according to the
scaling relation (\ref{KT_FSS_shifts}). From finite-lattice simulations the
polarization is determined by breaking the symmetry explicitly, i.e., if one defines
$\Sigma=\sum_i \sigma_i$, the spontaneous polarization is measured as
$P_0=\l|\Sigma|\r$ and the polarizability is estimated by
\begin{equation}
  \label{eq:susc_est}
  \chi = [\l \Sigma^2\r-\l |\Sigma|\r^2]/L^2.
\end{equation}
The peak locations of $\chi(\beta)$ were determined from simulations at nearby
couplings $\beta$ by means of the reweighting technique \cite{ferrenberg:88a}. The
phenomenological theory of FSS \cite{barber:domb} implies that the polarizability
$\chi$ for a finite lattice can be expressed as
\begin{equation}
  \label{eq:chi_fss_general}
  \chi(\beta,L) = L^{\gamma/\nu} X[L/\xi(\beta,\infty)],
\end{equation}
where $\xi(\beta,\infty)=\xi(\beta,L=\infty)$ denotes the correlation length of the
infinite system and $X$ is an analytic scaling function (here, we omit additional
irrelevant scaling fields representing corrections to scaling). Now, the maxima of
$\chi(\beta,L)$ correspond to the maximum of $X$ and thus all must occur at the same
value of the argument $L/\xi(\beta,\infty)$ (provided $X$ only has one maximum),
\begin{equation}
  \label{eq:fss_maxima}
  \frac{L}{\xi[\beta^\ast(L),\infty]} \equiv \kappa^{-1} = \const,
\end{equation}
thus defining a series of pseudo-critical temperatures $\beta^\ast(L)=\beta_\chi(L)$.
To find the general form of $\beta^\ast(L)$ in the scaling region, we need to solve
the expression (\ref{eq:corrlength}) for $\beta$. Inversion of the Taylor series of
$\xi$ of equation (\ref{eq:corrlength}) in powers of $y=\exp(-\pi^2/2\lambda)$ yields
\begin{equation}
  \label{eq:y_series}
  \lambda = -\frac{\pi^2}{2}\left(\ln\left[\frac{1}{4}\xi^{-1}-\frac{1}{48}\xi^{-3}+
    {\cal O}(\xi^{-5})\right]\right)^{-1},
\end{equation}
and $\beta(\lambda)$ expands around the critical point $\lambda=0$ as
\begin{equation}
  \label{eq:lambda_expansion}
  \beta = \ln 2 + \frac{1}{8}\lambda^2-\frac{1}{192}\lambda^4+{\cal O}(\lambda^6).
\end{equation}

\noindent
To leading order in both expansions one thus has via equation (\ref{eq:fss_maxima})
\begin{equation}
  \label{eq:fss_leading_order}
  \beta^\ast(L) = \beta_c + A_\beta(\ln 4\xi)^{-2} = \beta_c +A_\beta(\ln 4\kappa L)^{-2},
\end{equation}
where $A_\beta=(\pi^2/4\sqrt{2})^2$. Since in the FSS region $\xi\approx L$ and the
magnitude of the correction term $\xi^{-3}$ in equation (\ref{eq:y_series}) is
relatively suppressed by a factor of $10^{-4}$ already for the smallest lattice size
$L=16$ considered here, we conclude that this type of correction is not important at
the available level of statistical accuracy. Taking higher-order terms of
(\ref{eq:lambda_expansion}) into account, on the other hand, leads to the corrected
scaling form
\begin{equation}
  \label{eq:fss_corrected}
  \beta^\ast(L) = \beta_c+A_\beta(\ln 4\kappa L)^{-2}+C_\beta(\ln 4\kappa L)^{-4}
  + {\cal O}[(\ln 4\kappa L)^{-6}],
\end{equation}
with $C_\beta=(\pi^2/8\sqrt{3})^2$. We tested fits of this expected asymptotic form
to the simulation data for the toy model of an analytically generated series of
pseudo-critical points $\beta^\ast(L)$ defined by equation (\ref{eq:fss_maxima}) and
the exact form of the correlation length (\ref{eq:corrlength}) with $\kappa=1$ and
$L=16,24,\ldots,256$ as in the simulations, while taking $\beta_c$, $\kappa$ and the
amplitudes $A_\beta$ and $C_\beta$ as fit parameters. Already without the correction,
i.e., enforcing $C_\beta=0$, the critical coupling is reasonably reproduced as
$\beta_c=0.692$; the presence of neglected corrections shows up, however, in a fit
result $\kappa\approx 1.3$. Lifting the constraint on the amplitude $C_\beta$, one
arrives at $\beta_c=0.6931$ and $\kappa=1.05$, indicating perfect agreement with the
input data on the level of accuracy to be expected from the simulations.

Considering real simulation data, one might actually want to replace
$\xi(\beta,\infty)$ in (\ref{eq:fss_maxima}) by the finite-size expression
$\xi(\beta,L)$ (or, equivalently, allow the constant $\kappa$ to depend on system
size), introducing additional corrections not explicitly included in the scaling form
(\ref{eq:chi_fss_general})\footnote{Note that for the case of models with
  multiplicative logarithmic corrections, the replacement
  $\xi(\beta,\infty)\rightarrow\xi(\beta,L)$ on the r.h.s.\ of
  (\ref{eq:chi_fss_general}) has been suggested as the proper way of describing FSS
  in the first place \cite{kenna:04a}.}.  Exact expressions for the finite-size
behaviour of $\xi$ are not available\footnote{Note, however, that exact expression
  are available for the finite-size correlation length of the {\em XY\/} model on the
  strip geometry, cf.\ reference~\cite{balog01}.}, however, in view of the scaling
forms (\ref{fmodel_pscaling}) and (\ref{polariz_scaling_conj}) and the experience
with various models with logarithmic scaling corrections, it seems reasonable to make
the following general ansatz,
\begin{equation}
  \label{eq:xi_multiplicative_logarithmic}
  \xi[\beta^\ast(L),L] = \kappa L [1+A_\xi(\ln 4\kappa L)^{\omega_\xi}]
\end{equation}
with some {\em a priori\/} unknown exponent $\omega_\xi$. Depending on the sign of
$\omega_\xi$, this includes two basic inequivalent cases, namely a leading {\em
  multiplicative\/} logarithmic correction for $\omega_\xi>0$ or an {\em additive\/}
logarithmic correction for $\omega_\xi<0$. For the Ising and generalized $\phi^4$
models at their upper critical dimension, for instance, one has multiplicative
logarithmic corrections, corresponding to $\omega_\xi>0$, see, e.g.,
references~\cite{ballesteros:98b,bittner:02a,kenna:04a}. On the other hand, for the
two-dimensional $q=4$ Potts model \cite{salas:97a} as well as the two-dimensional
{\em XY\/} model \cite{hasenbusch:05a}, both of which are asymptotically related to
6-vertex models, only additive logarithmic corrections to the finite-size correlation
length occur at criticality. For positive $\omega_\xi$, the replacement
$\xi(\beta,\infty)\rightarrow\xi(\beta,L)$ in the derivation of the scaling of the
pseudo-critical temperatures $\beta^\ast(L)$ from (\ref{eq:fss_maxima}),
(\ref{eq:y_series}) and (\ref{eq:lambda_expansion}) produces a correction of the form
$\ln(\ln 4\kappa L)/(\ln 4\kappa L)^3$, whereas for integer $\omega_\xi < 0$ the
corrections can be expanded in a power series in $1/\ln4\kappa L$. To enable linear
fits, we finally express the scaling forms in terms of $L$ instead of $4\kappa L$,
leading to the following scaling descriptions:
 \begin{equation}
  \label{eq:final_effective_shifts}
  \beta^\ast(L) = \beta_c+A_\beta(\ln L)^{-2}+B_\beta(\ln L)^{-3} + C_\beta(\ln
  L)^{-4},\;\;\;\omega_\xi < 0,
\end{equation}
resp.\
\begin{equation}
  \label{eq:shifts_with_log_log_term}
  \beta^\ast(L) = \beta_c+A_\beta(\ln L)^{-2}\left[1+
    B_\beta\frac{\ln \ln L}{\ln L}\right], \;\;\;\omega_\xi>0.
\end{equation}
An indirect determination of the finite-size correlation length to be discussed below
in section \ref{sec_corr_scaling} strongly hints at the presence of only additive
logarithmic corrections in (\ref{eq:xi_multiplicative_logarithmic}), implying
$\omega_\xi < 0$, but in some cases both possibilities will be considered here to
illustrate the fact that a numerical discrimination between similar {\em forms\/} of
the corrections is not at all easily possible [note that due to the extremely slow
variation of the log-log term it might effectively be considered constant for the
range of lattice sizes considered, which would render the form
(\ref{eq:shifts_with_log_log_term}) equivalent to the ansatz
(\ref{eq:final_effective_shifts})].

\begin{figure}[tb]
  \centering
  \includegraphics[clip=true,keepaspectratio=true,width=9.5cm]{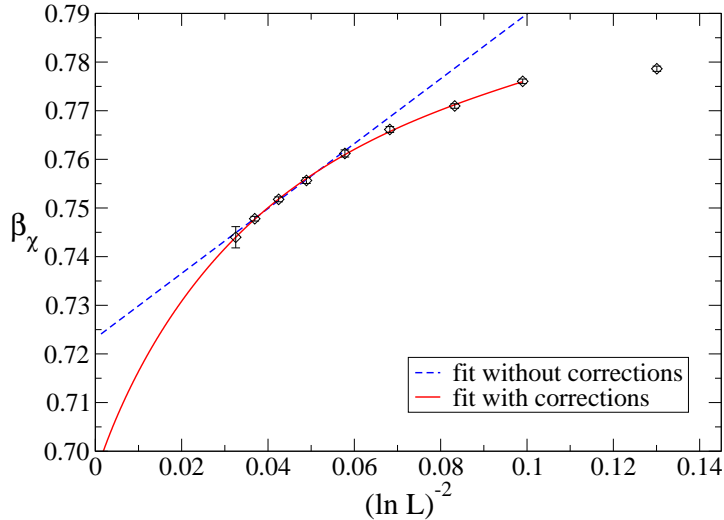}
  \caption
  {Peak positions of the staggered anti-ferroelectric polarizability of the $F$ model
    from simulations as a function of lattice size. The lines show fits of the form
    (\ref{eq:final_effective_shifts}) to the data. The dashed line corresponds to an
    uncorrected fit with $B_\beta=C_\beta=0$, starting from $L_\mathrm{min}=92$. The
    solid curve corresponds to a fit with corrections and $L_\mathrm{min}=32$.}
  \label{reg_F_peaks}
\end{figure}

\begin{table}[tb]
  \caption
  {Parameters of least-squares fits of the functional form
    (\ref{eq:final_effective_shifts}) to the simulation estimates for the peak locations
    of the staggered polarizability. No correction terms were taken into account,
    i.e., $B_\beta=C_\beta=0$ were held fixed throughout. $Q$ denotes the quality-of-fit
    parameter.\label{square_F_maxima_simple_table}}
  \begin{indented}
  \item[]\begin{tabular}{rr@{.}lr@{.}lr@{.}l} \br
      \multicolumn{1}{c}{$L_\mathrm{min}$} & \multicolumn{2}{c}{$\beta_c$} &
      \multicolumn{2}{c}{$A_\beta$} & \multicolumn{2}{c}{$Q$} \\ \mr
      16 & 0&73822(48) & 0&3547(62) & 0&00 \\
      24 & 0&73270(59) & 0&4533(87) & 0&00 \\
      32 & 0&73033(74) & 0&5018(126) & 0&00 \\
      46 & 0&72635(110) & 0&5912(223) & 0&46 \\
      64 & 0&72409(172) & 0&6453(385) & 0&88 \\
      92 & 0&72322(261) & 0&6668(624) & 0&78 \\
      128 & 0&72077(463) & 0&7307(1173) & 0&79 \\ \br
    \end{tabular}
  \end{indented}
\end{table}

The determined peak locations of the polarizability together with an example fit of
the functional form (\ref{eq:final_effective_shifts}) with omitted corrections, i.e.,
for $B_\beta=C_\beta=0$, to the data in the range $L=L_\mathrm{min}=92$ up to
$L=256$, are shown in figure \ref{reg_F_peaks}. The fit parameters of such fits,
successively omitting points from the low-$L$ side, are compiled in table
\ref{square_F_maxima_simple_table}. The strong deviations of the data from the form
with $B_\beta=C_\beta=0$ corresponding to a straight line in the chosen scaling of
the axes are apparent from figure \ref{reg_F_peaks}. Compared to the exact transition
point $\beta_c=\ln 2$, the estimates of $\beta_c$ from these fits are clearly too
large, dropping only very slowly as points from the small-$L$ side of the list are
successively omitted, cf.\ table \ref{square_F_maxima_simple_table}. One might
attempt to extrapolate these results towards $L_\mathrm{min}\rightarrow\infty$ using
the scaling form (\ref{eq:final_effective_shifts}); since the individual data are
highly correlated, however, this would introduce a strong bias. Instead, we directly
use the higher-order logarithmic corrections in the fitting procedure. Note that this
effect of strong scaling corrections here occurs for rather large lattices, where for
a usual continuous phase transition without logarithmic corrections the presence of
scaling corrections usually would not be much of an issue for the determination of
the leading scaling behaviour. Relaxing the constraints on $B_\beta$ only or on both
parameters, $B_\beta$ and $C_\beta$, we arrive at the fit results compiled in table
\ref{square_F_maxima_simple2_table}. It is apparent that the complexity of the
completely unconstrained fit type is at the verge of exceeding the available
statistical accuracy of the data, such that competing local minima of the $\chi^2$
distribution exist, which result in a rather discontinuous evolution of the
amplitudes $A_\beta$, $B_\beta$ and $C_\beta$ as the lower-end cut-off
$L_\mathrm{min}$ is increased. This functional form fits the data very well, however,
and the estimates for $\beta_c$ are all in agreement with the asymptotic value
$\beta_c=0.6931\ldots$ in terms of the statistical errors. It seems clear that the
remaining vague tendency of the fits to yield $\beta_c$ slightly above its asymptotic
value could in principle be removed by including further correction terms for the
case of extremely accurate data. Adding a $(\ln L)^{-5}$ term in
(\ref{eq:final_effective_shifts}), for instance, yields $\beta_c = 0.674(45)$,
$Q=0.72$ when including all data points.

\begin{table}[tb]
  \caption
  {Parameters of fits of the form (\ref{eq:final_effective_shifts}) to the peak
    locations of the polarizability. As indicated, $C_\beta=0$ was held fixed
    for the fits shown in the upper part of the table, while both parameters,
    $B_\beta$ and $C_\beta$, were allowed to vary in the fits presented in the lower part.
    \label{square_F_maxima_simple2_table}}
  \begin{indented}
  \item[]\begin{tabular}{rr@{.}lr@{.}lr@{.}lr@{.}lr@{.}l} \br
      \multicolumn{1}{c}{$L_{\mathrm{min}}$} & \multicolumn{2}{c}{$\beta_c$} &
      \multicolumn{2}{c}{$A_\beta$} & \multicolumn{2}{c}{$B_\beta$}&
      \multicolumn{2}{c}{$C_\beta$}& \multicolumn{2}{c}{$Q$} \\ \mr
      16 & 0&7103(17)  &    1&58(7)   & $-2$&91(17)  &    [0&0]     & 0&83\\
      24 & 0&7118(29)  &    1&50(14)  & $-2$&79(37)  &    [0&0]     & 0&78\\
      32 & 0&7069(46)  &    1&78(25)  & $-3$&47(67)  &    [0&0]     & 0&97\\
      46 & 0&7108(85)  &    1&53(51)  & $-2$&74(149) &    [0&0]     & 0&97\\
      64 & 0&714(16)   &    1&34(103) & $-2$&16(320) &    [0&0]     & 0&89\\ \mr
      16 & 0&7119(84)  &    1&44(71)  & $-2$&2(34)   &  $-0$&9(46)  & 0&73\\
      24 & 0&695(16)   &    3&2(16)   & $-11$&4(848) &    12&(124)  & 0&84\\
      32 & 0&723(32)   & $-0$&04(343) &    6&7(191)  & $-15$&8(297) & 0&96\\
      46 & 0&713(76)   &    1&3(93)   & $-1$&1(553)  &  $-2$&7(920) & 0&88\\ \br
    \end{tabular}
  \end{indented}
\end{table}

Using the alternative fit form (\ref{eq:shifts_with_log_log_term}) valid for
$\omega_\xi > 0$, on the other hand, results in fits quite similar to those obtained
from the ansatz (\ref{eq:final_effective_shifts}) with both correction terms present.
This relates back to the remark concerning the extremely slow variation of the log-log
term, which makes it plausible that the considered correction terms can be
effectively interchanged. No specific drift is observed on increasing the cutoff
$L_\mathrm{min}$. For $L_\mathrm{min}=64$ we arrive at $\beta_c = 0.695(40)$, $Q =
0.92$.  Hence, from the scaling of the pseudo-critical (inverse) temperatures
$\beta^\ast(L)$, a conclusion about the sign of $\omega_\xi$ can hardly be drawn.

\subsubsection{The correlation length.\label{sec_corr_scaling}}

\begin{figure}[tb]
  \centering
  \includegraphics[clip=true,keepaspectratio=true,width=9.5cm]{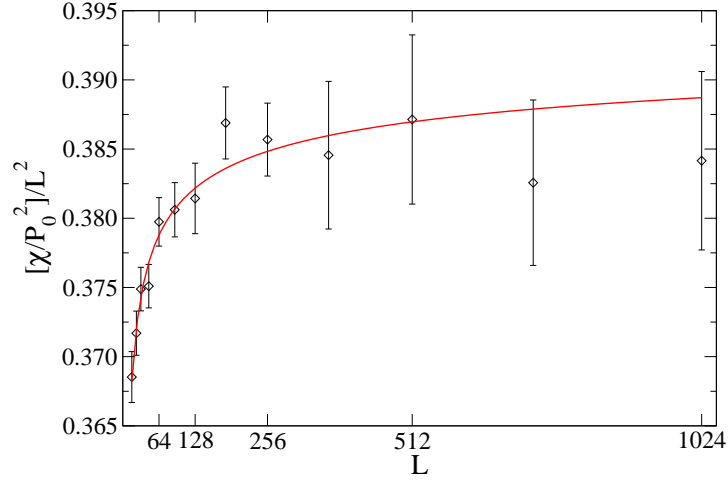}
  \caption
  {FSS of the combination $\chi(\beta, L)/P_0^2(\beta, L)$ at the asymptotic critical
    point $\beta_c=0.6931\ldots$ scaled by the leading $L^2$ behaviour to exhibit the
    corrections to scaling. The solid line shows a fit of the form
    (\ref{eq:xi_from_chiP0_with_corr}) to the data.}
  \label{fig:square_F_corr}
\end{figure}

Owing to the original relation of the present work to an investigation of the $F$
model on dynamical random graphs where the definition of connected correlation
functions is found to be highly non-trivial \cite{weigel:04b}, we have not measured
the correlation length directly. It turns out, however, that an indirect
determination is possible. Due to the scaling forms (\ref{fmodel_pscaling}) and
(\ref{polariz_scaling_conj}) of the polarization $P_0$ and the polarizability $\chi$,
for the combination $\chi/P_0^2$ the multiplicative logarithmic corrections cancel
such that to leading order
\begin{equation}
  \label{eq:chiP0}
  \chi(\beta, L)/P_0^2(\beta, L) = A_{\chi/P_0^2} \xi(\beta, L)^2
\end{equation}
in the scaling region, i.e., for $\beta=\beta^\ast(L)$. This relation allows for an
indirect determination of the exponent $\omega_\xi$ of the logarithmic correction of
the scaling form (\ref{eq:xi_multiplicative_logarithmic}),
\begin{equation}
  \label{eq:xi_from_chiP0}
  \chi(\beta^\ast, L)/P_0^2(\beta^\ast, L) = A_{\chi/P_0^2} L^2 [1+A_\xi(\ln L)^{\omega_\xi}]^2,
\end{equation}
where, again, the dependence on $\kappa$ has been dropped since its inclusion leads
to very badly converging, unstable fits. This corresponds to the omission (for the
time being) of higher-order corrections to scaling. From the simulation data, we find
the combination (\ref{eq:xi_from_chiP0}) at the critical point $\beta_c$ to be very
well described by a quadratic behaviour in $L$, the correction in square brackets
being quite small in absolute terms, cf.\ figure \ref{fig:square_F_corr}. Fitting the
form (\ref{eq:xi_from_chiP0}) to the data, we find clearly negative correction
exponents $\omega_\xi$ which, however, strongly depend on the cut-off
$L_\mathrm{min}$, systematically dropping from $\omega_\xi = -0.77(96)$ for
$L_\mathrm{min} = 16$ to, e.g., $\omega_\xi = -4.1(74)$ for $L_\mathrm{min} = 64$
with qualities $Q>0.8$. The large statistical errors on the estimate $\omega_\xi$
support the visual impression from figure \ref{fig:square_F_corr} that the correction
is actually too small to be reliably resolved at the present level of accuracy,
whereas the systematic drift results from the omission of higher-order corrections.
In fact, if we assume the familiar power-law form with negative exponents only,
\begin{equation}
  \label{eq:xi_from_chiP0_with_corr}
  \chi(\beta^\ast, L)/P_0^2(\beta^\ast, L) = A_{\chi/P_0^2} L^2 [1+A_\xi(\ln L)^{-1}+B_\xi(\ln
  L)^{-2}]^2,
\end{equation}
we find stable and high-quality (linear) fits for the amplitudes as $L_\mathrm{min}$
is varied, cf.\ the fit with $L_\mathrm{min} = 16$ and $Q=0.90$ displayed in figure
\ref{fig:square_F_corr}. Thus, in analogy to the cases of the two-dimensional {\em
  XY\/} \cite{hasenbusch:05a} and $q=4$ Potts \cite{salas:97a} models, the $F$ model
finite-size correlation length appears to exhibit only additive logarithmic
corrections corresponding to $\omega_\xi<0$ in
(\ref{eq:xi_multiplicative_logarithmic}).

\subsubsection{FSS of the spontaneous polarization.}

To derive FSS for the spontaneous polarization, consider the second form given in
equation (\ref{eq:polarization}),
$P_0^{1/2}=(2\pi/\lambda)^{1/2}[y^{1/4}+y^{9/4}+\ldots]$, which is rapidly convergent
in the scaling window. The sub-leading terms in $y$ are strongly suppressed in the
scaling regime and can be neglected compared to correction terms to follow. Using
again the expansion of $y$ in terms of $\xi=\xi(\beta,\infty)$ and equation
(\ref{eq:y_series}) for $\lambda(\xi)$, one arrives at
\begin{equation}
  \fl
  \label{eq:polarization_step1}
  P_0(\beta,\infty) = A_{P_0}(4\xi)^{-\beta/\nu}(\ln
  4\xi)^{\omega_{P_0}}\left[1-\frac{2}{3}(4\xi)^{-2}+
    \frac{4}{3}\frac{(4\xi)^{-2}}{\ln 4\xi}+\ldots\right],
\end{equation}
with $A_{P_0} = 4/\pi$, $\beta/\nu =1/2$ and $\omega_{P_0}=1$. Note that this (exact)
form does not contain any corrections of the log-log type present in the {\em XY\/}
model correlation function \cite{amitkadanoff}. To test the sufficiency of this
approximation, we again use an analytically generated, ``artificial'' series of
scaling data, evaluating $P_0(\beta,\infty)$ exactly from (\ref{eq:polarization}) for
the series of pseudo-critical temperatures defined by (\ref{eq:fss_maxima}) for
$\kappa=1$ and the exact expression (\ref{eq:corrlength}) for
$\xi=\xi(\beta,\infty)$. Fitting the form (\ref{eq:polarization_step1}) without the
corrections in square brackets to these data, taking $A_{P_0}$ and $\beta/\nu$ as fit
parameters (holding $\omega_{P_0}=1$ fixed), we arrive at $\beta/\nu = 0.5001$, which
is clearly sufficiently close to the exact result in terms of the statistical
accuracy to be expected from the simulation data. We thus conclude that the scaling
corrections in square brackets of (\ref{eq:polarization_step1}) can be neglected for
our purposes.

Further FSS corrections arise from the behaviour
(\ref{eq:xi_multiplicative_logarithmic}) of the finite-size correlation length. For
integer $\omega_\xi < 0$ as indicated by the investigation of $\chi/P_0^2$ above,
these corrections can be expanded in a power series,
\begin{equation}
  \label{eq:polarization_fss_series}
  \fl
  P_0(\beta^\ast,L) = A_{P_0} L^{-\beta/\nu} (\ln L)^{\omega_{P_0}}\left[1+\frac{B_{P_0}}{\ln
      L} + \frac{C_{P_0}}{(\ln L)^2}+\frac{D_{P_0}}{(\ln L)^3}\right],\;\;\;\omega_\xi < 0,
\end{equation}
where, again, the effect of the multiplier $\kappa$ is being incorporated in the
correction amplitudes. For the case of a positive $\omega_\xi$, on the other hand,
one would find a log-log correction to occur, i.e., one arrives at
\begin{equation}
  \label{eq:polarization_fss_with_loglog}
  \fl
  P_0(\beta^\ast,L) = A_{P_0} L^{-\beta/\nu}(\ln L)^{\omega_{P_0}-\omega_\xi/2}\left[
  1+B_{P_0}\frac{\ln\ln L}{\ln L}\right],\;\;\;\omega_\xi > 0
\end{equation}
where additionally, the exponent of the multiplicative logarithmic correction is
``dressed'' as $\omega_{P_0}-\omega_\xi/2$.

As for the route in the $(\beta^\ast(L),L)$ plane taken towards criticality, the two
principal choices are given by the determination of $P_0(\beta^\ast=\beta_c,L)$ at
the fixed asymptotic critical coupling $\beta_c=\ln 2$ {\em or\/} by considering
$P_0[\beta^\ast=\beta_\chi(L),L]$ at the polarizability peak locations
$\beta_\chi(L)$\footnote{Obviously, the first approach is only amenable in cases
  where $\beta_c$ is known {\em a priori\/}.}. Asymptotically, both approaches should
give compatible results; the strength and composition of scaling corrections,
however, might be noticeably different. Following the first approach, we analyze the
data from lattices of sizes $L=16,\ldots,1024$. Uncorrected fits of
(\ref{eq:polarization_fss_series}) with $B_{P_0} = C_{P_0} = D_{P_0} = 0$ and omitted
multiplicative correction term, i.e., $\omega_{P_0}=0$, yield exponents $\beta/\nu$
approaching the expected value logarithmically slow on successively omitting data
points from the small-$L$ side of the list. For $L=92,\ldots,1024$, for instance, we
find $\beta/\nu=0.4658(20)$, statistically incompatible with $\beta/\nu=1/2$. With
variable $\omega_{P_0}$, on the other hand, the leading scaling exponent can be
reasonably reproduced ($L_\mathrm{min}=24$),
\begin{equation}
  \begin{array}{rcl}
    A_{P_0} & = & 2.159(35), \\
    \beta/\nu & = & 0.4872(76), \\
    \omega_{P_0} & = & 0.109(33), \\
    Q & = & 0.14,
  \end{array}
\end{equation}
but the resulting exponent of the logarithmic correction is estimated in strong
deviation from $\omega_{P_0} = 1$. This shortcoming can only be remedied by including
the power-series type corrections in (\ref{eq:polarization_fss_series}). Letting only
$B_{P_0}$ vary, we arrive at estimates $\beta/\nu = 0.50(60)$, $\omega_{P_0} =
1.12(14)$, $Q = 0.13$, which fit the expectations very well. In view of the large
error estimates, however, one should not be deceived by the very small deviation from
the exact result. In fact, even with $C_{P_0} = D_{P_0} = 0$ still fixed, the
$\chi^2$ distribution exhibits multiple local minima and the fit results heavily
depend on the initial parameter values. Additionally letting $C_{P_0}$ and/or
$D_{P_0}$ vary, the fits get very unstable and meaningful results can no longer be
found. Using constraint fits, however, it can be clearly seen that the small value
$\omega_{P_0} = 0.109(33)$ above is indeed an effect of neglected higher-order
scaling corrections: fixing $\beta/\nu = 1/2$ as well as $\omega_{P_0}=1$ we
determine the amplitudes $A_{P_0}$, $B_{P_0}$, $C_{P_0}$ and $D_{P_0}$ with $Q=0.26$.
Now, fitting the form $E(\ln L)^{-\alpha}$ to the values of the thus determined
polynomial $1+B_{P_0}(\ln L)^{-1}+C_{P_0}(\ln L)^{-2}+D_{P_0}(\ln L)^{-3}$, we find
$\alpha\approx 0.83$. Thus, neglecting the scaling corrections in square brackets of
(\ref{eq:polarization_fss_series}) clearly leads to an effective reduction of the
exponent estimate $\omega_{P_0}$ from its asymptotic value $\omega_{P_0} = 1$ by
about $\alpha \approx 0.8-0.9$.

Considering the spontaneous polarization at the peak positions of the polarizability
for lattice sizes up to $L=256$, the uncorrected form with $\omega_{P_0} = B_{P_0} =
C_{P_0} = D_{P_0} = 0$ yields very small estimates for $\beta/\nu$ around
$\beta/\nu\approx 0.25$ slowly increasing with the cutoff $L_\mathrm{min}$.
Including the multiplicative logarithmic correction of equation
(\ref{eq:polarization_fss_series}), i.e., relaxing the constraint $\omega_{P_0}=0$,
these results can be improved, and, e.g., for $L_\mathrm{min}=92$, we find the
following fit parameters,
\begin{equation}
  \begin{array}{rcl}
    A_{P_0} & = & 1.49(50), \\
    \beta/\nu & = & 0.44(11), \\
    \omega_{P_0} & = & 0.71(55), \\
    Q & = & 0.57,
  \end{array}
\end{equation}
with an exponent estimate for $\beta/\nu$ well compatible with the exact result
$\beta/\nu=1/2$, although endowed with an unpleasantly large statistical error. The
inclusion of the power-series type scaling corrections of
(\ref{eq:polarization_fss_series}) necessary for a full resolution of the corrections
is not possible with the present data. Fixing $\beta/\nu = 1/2$, $\omega_{P_0} = 1$,
however, these terms provide an excellent description for the present scaling
corrections and a quality-of-fit $Q=0.72$ is already attained for $L_\mathrm{min} =
16$ (including all three terms $B_{P_0}$, $C_{P_0}$ and $D_{P_0}$). In general, fits
to the data at the maxima of the polarizability are found to be somewhat less stable
and precise than those to the data at $\beta_c$, which we attribute to the smaller
available system sizes here as well as additional scatter of the data due to the
necessary reweighting.

In addition to the self-contained scaling routes in the $(\beta^\ast(L),L)$ plane
described, for the exactly solved case considered here it is also possible to perform
simulations for the analytical series $\beta^\ast(L)$ of inverse temperatures defined
by the relation (\ref{eq:fss_maxima}) with the exact expression (\ref{eq:corrlength})
for $\xi$, which yields a scenario somewhat in between the $\beta^\ast(L)=\beta_c$
and $\beta^\ast(L)=\beta_\chi(L)$ cases. This artificial series of simulation data is
indeed found to result in quite stable fits, such that at least the amplitude
$B_{P_0}$ of (\ref{eq:polarization_fss_series}) can be left variable to yield
$\beta/\nu=0.51(47)$, $\omega_{P_0}=1.1(69)$ with $Q=0.66$ and $L_\mathrm{min}=16$.
It should be noted that also fits of the form (\ref{eq:polarization_fss_with_loglog})
with a log-log correction are possible with good quality for the scaling of the
polarization, which only very slightly change the estimates for $\beta/\nu$, but do
not lift the estimate for $\omega_{P_0}$ to the expected value $\omega_{P_0} = 1$.
Thus, it would be hard to distinguish the forms (\ref{eq:polarization_fss_series})
and (\ref{eq:polarization_fss_with_loglog}) solely on the basis of the numerical
polarization data.

\subsubsection{FSS of the polarizability.}

\begin{figure}[tb]  
  \centering
  \includegraphics[clip=true,keepaspectratio=true,width=9.5cm]{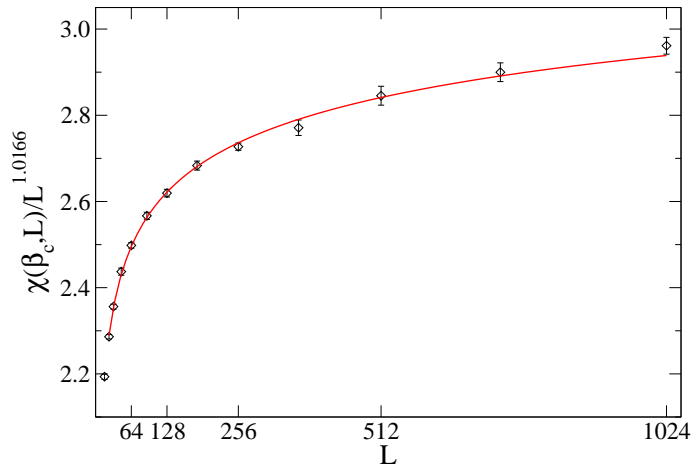}
  \caption
  {Finite-size scaling plot of the critical staggered polarizability
    $\chi(\beta_c,L)$ for lattice sizes from $L=16$ up to $L=1024$. The solid line
    shows a fit of the functional form (\ref{eq:polarizability_fss_form}) with
    $B_\chi=C_\chi=D_\chi=0$ to the data. The abscissa has been re-scaled such as to
    factor out the leading scaling behaviour $\propto L^{\gamma/\nu}$ with
    $\gamma/\nu=1.0166$ from the fit (\ref{eq:gamma_critical_fit_result}).}
  \label{reg_F_chi}
\end{figure}

Since the staggered polarizability $\chi$ is not known exactly, a systematic
discussion of $\chi$ as a function of the asymptotic correlation length $\xi$ is not
possible. However, from Baxter's conjecture (\ref{polariz_scaling_conj}) the leading
behaviour is expected to be
\begin{equation}
  \label{eq:polarizability_starting_point}
  \chi(\beta,\infty) = A_\chi \xi^{\gamma/\nu}(\ln \xi)^{\omega_\chi},
\end{equation}
with $\gamma/\nu = 1$ and $\omega_\chi = 2$. Repeating the arguments presented above
for the polarization, again assuming an integer $\omega_\xi < 0$ in equation
(\ref{eq:xi_multiplicative_logarithmic}), one deduces the following FSS ansatz,
\begin{equation}
  \label{eq:polarizability_fss_form}
  \fl 
  \chi(\beta^\ast,L) = A_\chi L^{\gamma/\nu} (\ln L)^{\omega_\chi}\left[1+\frac{B_\chi}{\ln
    L} + \frac{C_\chi}{(\ln L)^2}+\frac{D_\chi}{(\ln L)^3}\right],\;\;\;\omega_\xi < 0,
\end{equation}
whereas $\omega_\xi > 0$ would result in a form including a log-log correction as in
(\ref{eq:polarization_fss_with_loglog}). We first investigate the simulation results
at criticality, using data from lattices of sizes $L=16,\ldots,1024$. From fits of
the leading scaling behaviour dropping the multiplicative and additive logarithmic
correction terms, $\omega_\chi=0$ and $B_\chi=C_\chi=D_\chi=0$, to these data, we
find reasonable fit qualities only when dropping many points from the small-$L$ side
of the size range.  Successively increasing the cutoff $L_\mathrm{min}$ a very slow
downward drift of the estimates for $\gamma/\nu$ is observed. For
$L_\mathrm{min}=92$, we arrive at an estimate $\gamma/\nu = 1.0754(22)$ with
$Q=0.96$, which is clearly incompatible with the exact result in terms of the
statistical error. Letting $\omega_\chi$ vary while still keeping
$B_\chi=C_\chi=D_\chi=0$ fixed, stable and good-quality fits can be attained.
For the range $L=24,\ldots,1024$, we have
\begin{equation}
  \begin{array}{rcl}
    A_\chi & = & 1.581(31), \\
    \gamma/\nu & = & 1.0166(90), \\
    \omega_\chi & = & 0.320(40), \\
    Q & = & 0.78,
  \end{array}
  \label{eq:gamma_critical_fit_result}
\end{equation}
in good agreement with the exact result $\gamma/\nu=1$, however, again clearly
missing the expected asymptotic value of the exponent of the multiplicative
logarithmic correction, $\omega_\chi = 2$. The scaling plot presented in figure
\ref{reg_F_chi} shows this last fit together with the simulation data, scaled such as
to expose the magnitude of scaling corrections present. The asymptotic value
$\omega_\chi = 2$ could be recovered by including the correction amplitudes $B_\chi$,
$C_\chi$ and $D_\chi$. Letting only $B_\chi$ additionally vary, $\omega_\chi$ is
already increased to $\omega_\chi = 1.32(17)$ with $\gamma/\nu = 1.0(48)$, $Q = 0.78$
and $L_\mathrm{min} = 16$. The obviously necessary higher-order terms $C_\chi$ and
$D_\chi$ unfortunately cannot be fitted any more with the present data, however. On
the other hand, fixing again $\gamma/\nu = 1$ and $\omega_\chi = 2$, the present
corrections can be well described by the amplitudes $B_\chi$, $C_\chi$, $D_\chi$,
resulting in a quality-of-fit of $Q=0.84$ already for $L_\mathrm{min} = 16$. We note
that here even the inclusion of the $(\ln L)^{-3}$ term $D_\chi$ is probably crucial
since the leading multiplicative logarithmic correction is already quadratic.

Estimates of the maxima $\chi[\beta_\chi(L),L]$ are available for lattice sizes
$L=16,\ldots,256$. A reasonable quality fit of the uncorrected form
(\ref{eq:polarizability_fss_form}) with $\omega_\chi=0$ and $B_\chi=C_\chi=D_\chi=0$
to these data can be produced starting from $L_\mathrm{min}=64$, which yields an
estimate $\gamma/\nu=1.2788(58)$, $Q=0.23$, lying even much further off the
asymptotic result than in the case of the critical polarizability. Letting
$\omega_\chi$ vary while keeping $B_\chi=C_\chi=D_\chi=0$ fixed, the estimate for
$\gamma/\nu$ is noticeably reduced to $\gamma/\nu=1.13(08)$ with
$\omega_\chi=0.71(36)$, $Q=0.15$, for $L_\mathrm{min}=46$, and a further tendency to
decrease on an increase of $L_\mathrm{min}$ remains. Although here again, the need
for higher-order correction terms is apparent, we find the data not precise enough
for their inclusion. Thus, although both methods, consideration of the critical
polarizability as well as scaling of the peak heights of $\chi(L)$, yield equivalent
results, we find corrections to scaling slightly more pronounced in the latter
approach. This is partly explained by the fact that for $\chi(\beta_c,L)$ larger
lattice sizes could be considered. However, even restricting a fit with
$B_\chi=C_\chi=D_\chi=0$ for $\chi(\beta_c,L)$ to $L\le256$, we find with
$\gamma/\nu=1.006(11)$ for $L_\mathrm{min}=16$ a considerably more precise result
closer to the asymptotic value; additionally, as mentioned above, no further drift of
$\gamma/\nu$ is noticeable there as $L_\mathrm{min}$ is increased.  For the extra
simulation series at inverse temperatures $\beta^\ast(L)$ resulting from equation
(\ref{eq:fss_maxima}) with $\kappa=1$, a fit with $B_\chi=C_\chi=D_\chi=0$ and
$L_\mathrm{min}=32$ leads to $\gamma/\nu = 1.075(66)$, $\omega_\chi = 1.21(29)$ with
$Q=0.48$. Inclusion of the $B_\chi$, $C_\chi$ and $D_\chi$ terms destabilizes the
fits too far, although consistency with $\gamma/\nu = 1$, $\omega_\chi = 2$ is again
found on fitting the amplitudes only.

\subsubsection{The scaling dimension in the critical phase.}

\begin{figure}[tb]  
  \centering
  \psfrag{this is fit number one or somethin}{\small $x_P$ from fits to
    (\ref{eq:corrected_critical_phase_scaling})}
  \psfrag{this is fit number two}{\small $x_P$ from fits to
    (\ref{eq:leading_critical_phase_scaling})}
  \psfrag{this is fit number three}{\small $2-2x_P$ from fits to
    (\ref{eq:leading_critical_phase_scaling})}
  \psfrag{this is fit number four}{\small $2-2x_P$ from fits to
    (\ref{eq:corrected_critical_phase_scaling})}
  \includegraphics[clip=true,keepaspectratio=true,width=9.5cm]{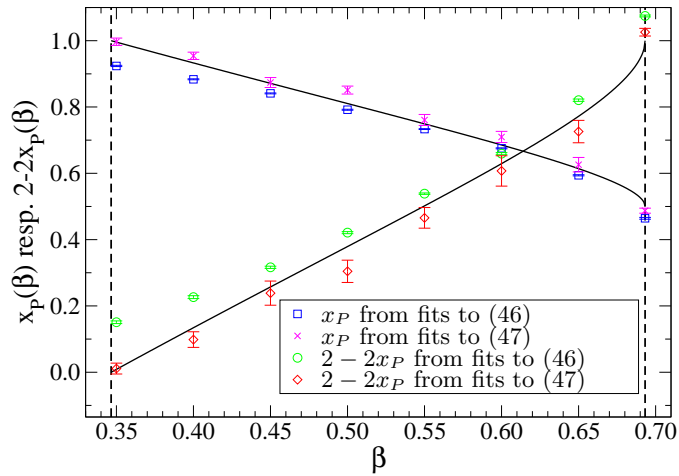}
  \caption
  {Finite-size scaling dimensions $x_P(\beta)$ of the spontaneous staggered
    polarization and $2-2x_P(\beta)$ of the staggered polarizability, respectively,
    as a function of inverse temperature $\beta$ in the critical phase
    $\beta<\beta_c$. The symbols denote results from FSS fits of the functional forms
    (\ref{eq:leading_critical_phase_scaling}) resp.\
    (\ref{eq:corrected_critical_phase_scaling}) with $B=C=D=0$ to the simulation
    data. The solid lines correspond to the conjecture
    (\ref{eq:rewritten_critical_scaling}) for the analytic form.  The vertical dashed
    lines indicate the locations of the the free-fermion point
    $\beta_f=\frac{1}{2}\ln 2$ and the critical point $\beta_c=\ln 2$, respectively.}
  \label{fig:critical_gamma}
\end{figure}

Due to the criticality of the high-temperature phase one expects scaling and,
accordingly, FSS in the whole region $\beta<\beta_c=\ln 2$. The closed-form
conjecture (\ref{eq:delta_conjecture}) for the exponent $\eta$ entails predictions
for the FSS of $P_0(\beta,L)$ and $\chi(\beta,L)$ for $\beta<\beta_c$. In terms of
the scaling dimension $x_P=\beta/\nu=1-\gamma/2\nu$ \cite{henkel:book} and the
inverse temperature $\beta$, equation (\ref{eq:delta_conjecture}) reads
\begin{equation}
  x_P(\beta<\beta_c)=\frac{\pi}{2}\left\{\mathrm{arccos}[1-\frac{1}{2}\exp(2\beta)]\right\}^{-1},
  \label{eq:rewritten_critical_scaling}
\end{equation}
which behaves close to the critical point $\beta_c=\ln 2$ as
\begin{equation}
  \label{eq:xp_expansion}
  x_P(\beta)=\frac{1}{2}+\frac{\sqrt{2}}{\pi}(\ln2-\beta)^{1/2}+{\cal O}(\ln2-\beta),
\end{equation}
such that $x_P$ has a vertical tangent at $\beta_c$, implying an especially sensitive
dependence of $x_P$ on scaling corrections there. It is worthwhile to notice that,
although the correspondence between the {\em XY\/} and $F$ models only applies to their
critical points, an analogous square-root singularity of the exponent $\eta$ of the
{\em XY\/} model is found on entering the critical low-temperature phase there, see, e.g.,
reference \cite{res:00a}.  The leading scaling behaviour of $P_0(\beta,L)$ and
$\chi(\beta,L)$ for $\beta<\beta_c$ is hence expected to be
\begin{equation}
  \label{eq:leading_critical_phase_scaling}
  \begin{array}{rcl}
    P_0(\beta,L) & = & A_{P_0}L^{-x_P(\beta)},\\
    \chi(\beta,L) & = & A_\chi L^{2-2x_P(\beta)}.
  \end{array}
\end{equation}
The solid lines of figure \ref{fig:critical_gamma} illustrate the predicted behaviour
of these exponents in the high-temperature phase. As can be seen, the polarizability
exponent $2-2x_P(\beta)$ crosses zero at the free-fermion coupling
$\beta_f=\frac{1}{2}\ln 2$ and, consequently, $\chi$ should be non-divergent below.
As a result, the predicted singularity would be covered by non-singular background
terms there, such that we restrict ourselves to the range $\beta_f\le \beta\le
\beta_c$ here.

\begin{table}[tb]
  \small
  \setlength{\tabcolsep}{0.85ex}
  \caption{Fit parameters for $P_0$ and $\chi$ in the critical phase
    $\beta_f\le\beta\le\beta_c$. (a) Fits of the form
    (\ref{eq:leading_critical_phase_scaling}). (b) Fits of the form
    (\ref{eq:corrected_critical_phase_scaling}) with $B=C=D=0$. (c) Fits of the form
    (\ref{eq:corrected_critical_phase_scaling}) with $D=0$ and $x_P$ fixed at the
    values (\ref{eq:rewritten_critical_scaling}). (d) Fits of the form
    (\ref{eq:corrected_critical_phase_scaling}) with $D=0$ and $\omega_{P_0}=1$
    resp.\ $\omega_\chi=2$ fixed.\label{tab:critical_phase_fits}}
  \begin{tabular}{lr@{.}lr@{.}llr@{.}lr@{.}llr@{.}lr@{.}lr@{.}lr@{.}l}\br
    & \centre{2}{conj.} & \centre{3}{(a)} &
    \centre{5}{(b)} & \centre{4}{(c)} & \centre{4}{(d)} \\ \ns\ns
    & \crule{2} & \crule{3} & \crule{5} & \crule{4} & \crule{4} \\ 
    \centre{1}{$\beta$} & \centre{2}{$x_P$} & \centre{2}{$x_P$} & \centre{1}{$L_\mathrm{min}$} &
    \centre{2}{$x_P$} & \centre{2}{$\omega_{P_0}$} & \centre{1}{$L_\mathrm{min}$} &
    \centre{2}{$\omega_{P_0}$} & \centre{2}{$Q$} & \centre{2}{$x_P$} & \centre{2}{$Q$} \\ \mr
    $\ln 2$ & 0&500 & 0&4643(27) & 128 &  0&487(08) & 0&109(33) & 24 & 1&16(32) & 0&11 & 0&48(62) & 0&20\\
    0.65    & 0&614 & 0&5940(18) & 128 &  0&626(22) & 0&18(12)  & 92 & 1&01(18) & 0&47 & 0&61(44) & 0&44\\
    0.60    & 0&685 & 0&6753(15) & 182 &  0&710(17) & 0&212(96) & 92 & 0&95(15) & 0&61 & 0&69(38) & 0&67\\
    0.55    & 0&749 & 0&7334(12) & 128 &  0&761(17) & 0&161(94) & 92 & 1&00(17) & 0&45 & 0&75(40) & 0&45\\
    0.50    & 0&811 & 0&7917(15) & 182 &  0&851(12) & 0&354(66) & 92 & 1&04(15) & 0&59 & 0&81(36) & 0&49\\
    0.45    & 0&871 & 0&8411(14) & 182 &  0&874(15) & 0&206(85) & 92 & 1&11(15) & 0&02 & 0&86(34) & 0&01\\
    0.40    & 0&933 & 0&8835(14) & 182 &  0&954(11) & 0&424(60) & 64 & 1&26(19) & 0&55 & 0&91(25) & 0&10\\
    0.35    & 0&996 & 0&9238(18) & 256 &  0&997(11) & 0&456(61) & 64 & 1&46(22) & 0&74 & 0&95(21) & 0&07\\ \mr
    \centre{1}{$\beta$} & \centre{2}{$2-2x_P$} & \centre{2}{$2-2x_P$} &
    \centre{1}{$L_\mathrm{min}$} & \centre{2}{$2-2x_P$} &
    \centre{2}{$\omega_\chi'$} & \centre{1}{$L_\mathrm{min}$} &
    \centre{2}{$\omega_\chi$} & \centre{2}{$Q$} & \centre{2}{$2-2x_P$} & \centre{2}{$Q$} \\ \mr
    $\ln 2$ & 1&000 & 1&0746(28) & 128 &  1&017(09) & 0&32(04) & 24   & 2&50(12) & 0&91 & 0&96(309)  & 0&53\\
    0.65    & 0&772 & 0&8206(35) & 182 &  0&726(34) & 0&56(19) & 92   & 2&81(11) & 0&24 & 0&66(320)  & 0&26\\
    0.60    & 0&629 & 0&6593(32) & 182 &  0&607(46) & 0&32(26) & 128  & 2&85(08) & 0&99 & 0&51(640)  & 0&45\\
    0.55    & 0&502 & 0&5386(21) & 128 &  0&466(31) & 0&42(18) & 92   & 2&80(08) & 0&79 & 0&39(668)  & 0&11\\
    0.50    & 0&379 & 0&4211(31) & 182 &  0&304(34) & 0&70(19) & 92   & 2&64(14) & 0&75 & 0&28(825)  & 0&73\\
    0.45    & 0&257 & 0&3165(33) & 182 &  0&239(36) & 0&50(21) & 92   & 2&93(10) & 0&03 & 0&17(1132) & 0&01\\
    0.40    & 0&134 & 0&2265(44) & 256 &  0&098(23) & 0&81(13) & 64   & 2&81(20) & 0&83 & 0&08(1362) & 0&85\\
    0.35    & 0&009 & 0&1508(40) & 256 &  0&011(16) & 0&88(09) & 46   & 2&85(26) & 0&56 & 0&00(1478) & 0&64\\ \br
  \end{tabular}
\end{table}

To test the form (\ref{eq:rewritten_critical_scaling}) we performed seven series of
simulations at inverse temperatures $\beta=0.35,0.40,\ldots,0.65$ with the same
series of system sizes $L=16,\ldots,1024$ used at $\beta=\ln 2$. Fitting the expected
leading scaling behaviour (\ref{eq:leading_critical_phase_scaling}) to the simulation
data, many system sizes from the small-$L$ side have to be dropped to reach
satisfactory fit qualities and to account for the observed slow drift of the
resulting scaling exponents on increasing $L_\mathrm{min}$, which was finally chosen
to be $L_\mathrm{min}=182$ in most cases, cf.\ the data in column (a) of table
\ref{tab:critical_phase_fits}. As can be seen from the fit data presented in figure
\ref{fig:critical_gamma}, even with this precaution highly significant deviations of
the fit results from equation (\ref{eq:rewritten_critical_scaling}) are observed,
especially close to the free-fermion coupling $\beta_f$. Scaling corrections are
assumed here to take the form found at criticality, i.e.,
\begin{equation}
  \label{eq:corrected_critical_phase_scaling}
  \fl
  \begin{array}{rcl}
    P_0(\beta,L) & = & \ds A_{P_0}L^{-x_P(\beta)}(\ln L)^{\omega_{P_0}(\beta)}\left[1+
      \frac{B_{P_0}}{\ln L} + \frac{C_{P_0}}{(\ln L)^2}+\frac{D_{P_0}}{(\ln L)^3}
    \right],\\[1.75ex]
    \chi(\beta,L) & = & \ds A_\chi L^{2-2x_P(\beta)}(\ln L)^{\omega_{\chi}(\beta)}\left[1+
      \frac{B_\chi}{\ln L} + \frac{C_\chi}{(\ln L)^2}+\frac{D_\chi}{(\ln L)^3}
      \right].
  \end{array}
\end{equation}
As for the critical polarization and polarizability, fits including all of the
correction terms (amounting to six variable parameters) are not possible with the
available data. Including only the multiplicative logarithmic correction with
variable exponent $\omega_{P_0}$ resp.\ $\omega_\chi$, we arrive at largely improved
estimates for the scaling dimension $x_P$ in agreement with the prediction
(\ref{eq:rewritten_critical_scaling}), see the column (b) of table
\ref{tab:critical_phase_fits} and the data in figure \ref{fig:critical_gamma}. The
values of the correction exponents $\omega_{P_0}$ resp.\ $\omega_\chi$, however,
again have to be considered as effective exponents owed to the omission of the
additive corrections $B$, $C$, $D$. Note that in principle also the values of
$\omega_{P_0}$ resp.\ $\omega_\chi$ could depend on the value of the coupling $\beta$
as indicated in equation (\ref{eq:corrected_critical_phase_scaling}). To investigate
this possibility, we performed fits with the leading scaling exponents fixed to the
presumably exact values of $x_P$ from (\ref{eq:rewritten_critical_scaling}), letting
$\omega_{P_0}$ resp.\ $\omega_\chi$ vary and including two orders of additive scaling
corrections, i.e., enforcing $D=0$ only. The results of these fits are collected in
column (c) of table \ref{tab:critical_phase_fits}. In all cases with the exception of
$\beta = 0.45$, which seems to be an outlier, we find very good fit qualities, again
indicating consistency with the conjecture (\ref{eq:rewritten_critical_scaling}). The
estimates for $\omega_{P_0}$ are all consistent with a constant value of
$\omega_{P_0} = 1$, independent of the coupling $\beta$. The estimates for
$\omega_\chi$, on the other hand, are clearly larger than $\omega_\chi = 2$, but no
general trend on varying $\beta$ is observed. This deviation of $\omega_\chi$ is
found to disappear upon inclusion of the next-order correction amplitude $D_\chi$ for
which, however, both exponents, $x_P$ and $\omega_\chi$, have to be kept fixed. This
term has to be included here but not for the polarization since for $\chi$ already
the leading multiplicative logarithmic correction is quadratic. When fixing both
exponents, $x_P$ and $\omega$, fits of good quality are attained for both observables
and all couplings $\beta$ on including all three additive correction terms of
(\ref{eq:corrected_critical_phase_scaling}). In passing we note that an analysis of
$\chi/P_0^2$ in the critical phase yields negative values of $\omega_\xi$ everywhere
and fits of the power-law form (\ref{eq:xi_from_chiP0_with_corr}) describe the
corrections extremely well.

\subsection{Results of the Thermal Scaling Analysis}

\begin{figure}[tb]  
  \centering
  \includegraphics[clip=true,keepaspectratio=true,width=9.5cm]{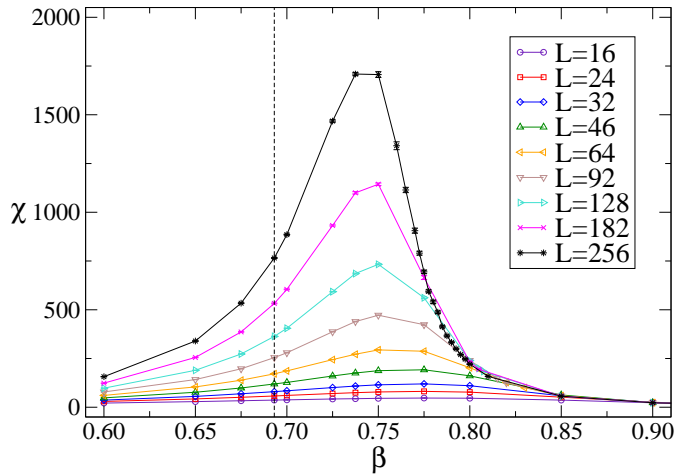}
  \caption
  {Scaling of the polarizability peaks from simulation data. The lines simply connect
    the data points and are drawn to guide the eye. The dashed vertical line
    indicates the location of the asymptotic critical coupling
    $\beta_c=0.6931\ldots$}
  \label{reg_F_overview}
\end{figure}

The discussed FSS of the critical polarization and polarizability is independent of
the value of the critical exponent $\rho$. For the scaling of the polarizability peak
positions in section \ref{sec:critical_coupling}, on the other hand, the need to
resolve the present strong logarithmic scaling corrections did not allow for an
additional independent determination of $\rho$. To directly verify the exponential
type of the observed divergences and to estimate the parameter $\rho$, one should
hence consider {\em thermal\/} instead of finite-size scaling. Figure
\ref{reg_F_overview} shows an overview of the temperature dependence of the staggered
polarizability for different lattice sizes. The clear scaling of $\chi$ for the
high-temperature region $\beta<\beta_c=\ln 2$ illustrates again the presence of a
critical phase. In contrast, for the low-temperature phase to the right of the peaks,
the polarizability curves essentially collapse and only start to disagree as the
correlation length reaches the linear extent of the considered lattice.  Therefore, a
thermal scaling analysis must be performed in the low-temperature vicinity of the
critical point, the behaviour in the high-temperature phase being completely governed
by finite-size effects. Here, we do not consider the scaling of the correlation
length itself, but instead analyze the thermal scaling of the spontaneous
polarization and the polarizability for a single lattice of size $L=256$.
Simulations were performed for a closely spaced series of temperatures in the
low-temperature vicinity of the critical point.

\subsubsection{Scaling of the spontaneous polarization.}

\begin{figure}[tb]  
  \centering
  \includegraphics[clip=true,keepaspectratio=true,width=9.5cm]{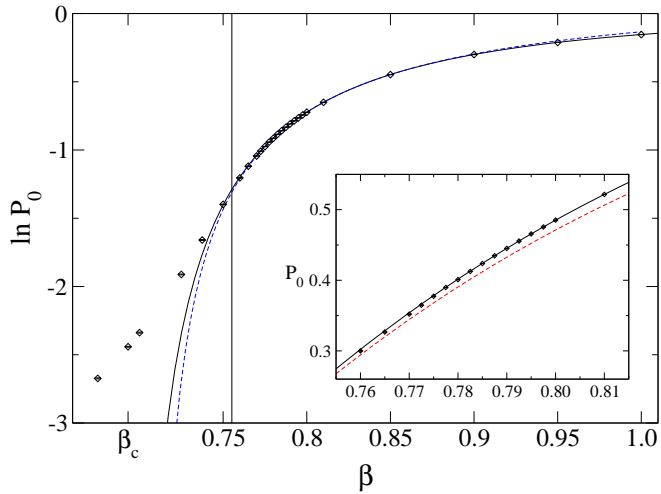}
  \caption
  {Thermal behaviour of the spontaneous staggered polarization $P_0$ close to
    $\beta_c$ from simulations of a $256^2$ system. The solid line denotes the exact
    asymptotic result (\ref{eq:polarization}), the dashed line is a fit of the form
    (\ref{eq:polarization_thermal_scaling}) to the data. The vertical line denotes
    the point where $\xi(\beta^\ast,\infty)=L=256$. The inset shows the exact
    solution (\ref{eq:polarization}) compared to the first-order approximation of
    (\ref{eq:polarization_expansion}) in the inverse temperature regime used for the
    fit.}
  \label{square_F_P_thermal}
\end{figure}

From the leading term of (\ref{eq:polarization}) in $y$ and the dependence of
$\lambda$ on $\beta$, the spontaneous polarization behaves as
\begin{eqnarray}
  \label{eq:polarization_expansion}
  P_0(\beta) & = & 
  \frac{\pi}{\sqrt{2}}\left[(\beta-\beta_c)^{-1/2}-\frac{1}{6}(\beta-\beta_c)^{1/2}
  +\ldots\right]  \nonumber \\
  & & \times\exp\left\{-\frac{\pi^2}{8\sqrt{2}}\left[(\beta-\beta_c)^{-1/2}-
        \frac{1}{6}(\beta-\beta_c)^{1/2}+\ldots\right]\right\}
\end{eqnarray}
as $\beta_c$ is approached from above. Taking only the leading-order terms into
account, we consider the following scaling form,
\begin{equation}
  \label{eq:polarization_thermal_scaling}
  \ln P_0(\beta) = A_{P_0}+B_{P_0}(\beta-\beta_c)^{-\rho}+C_{P_0}\ln(\beta-\beta_c),
\end{equation}
with $C_{P_0}=-1/2$ and $\rho=1/2$. The window of validity of
(\ref{eq:polarization_thermal_scaling}) for the thermal scaling of $P_0$ for a finite
lattice is limited for small deviations $\beta-\beta_c$ by finite-size effects and
for large deviations $\beta-\beta_c$ by the higher-order corrections to scaling
indicated in (\ref{eq:polarization_expansion}). If correlation lengths are measured,
one might monitor the effect of the finite lattice size by comparing the value of the
correlation length $\xi(\beta, L)$ at a given $\beta>\beta_c$ with the linear extent
$L$ of the lattice \cite{wj:93a}\footnote{Although the behaviour of the finite-size
  correlation length has been indirectly analyzed above in section
  \ref{sec_corr_scaling}, unfortunately we do not have access to the amplitude to
  find the absolute values of $\xi(\beta,L)$.}. Here, the onset of finite-size
effects is estimated by the beginning of the rounding of the exponential decline of
$P_0$ as $\beta_c$ is approached. From monitoring the quality-of-fit parameter and
estimation of the onset of the finite-size rounding, we determine a fit range of
$\beta_\mathrm{min}=0.77\le\beta\le 0.85=\beta_\mathrm{max}$. We find fits of the
full five-parameter family (\ref{eq:polarization_thermal_scaling}) of functions to
the data possible, but the resulting fit parameters are endowed with astronomic error
estimates and the corresponding $\chi^2$ distribution has multiple minima such that
different ``solutions'' can be found. We thus fix one or two of the parameters at
their expected asymptotic values to reach more stable fits, cf.\ the fit data
collected in table \ref{tab:thermal_fits}. Note that the parameters of the fully
unrestricted fit where found starting from the parameters of one of the restricted
fits, thus explicitly selecting one of the $\chi^2$ minima. Figure
\ref{square_F_P_thermal} shows the simulation data together with this unrestricted
fit and the exact asymptotic polarization of (\ref{eq:polarization}). The vertical
line denotes the inverse temperature $\beta^\ast$ where the asymptotic correlation
length $\xi(\beta^\ast,\infty)$ of equation (\ref{eq:corrlength}) reaches the linear
size $L=256$ of the system. As expected, this point approximately coincides with the
inverse temperature where the simulation data deviate from the asymptotic result due
to finite-size effects, thus justifying the method of determining
$\beta_\mathrm{min}$. The inset of figure \ref{square_F_P_thermal} shows the
approximation of (\ref{eq:polarization_expansion}) with only the first-order terms of
both expansions being kept in comparison to the full asymptotic result
(\ref{eq:polarization}) and the simulation data. As can be seen, even in the scaling
range considered here, the deviation is much larger than the statistical errors of
the data. The observed shift, however, can be mostly reproduced by slight changes of
the amplitudes $A_{P_0}$ and $B_{P_0}$, such that
(\ref{eq:polarization_thermal_scaling}) still fits the data well. The fitted
amplitudes $A_{P_0}$, $B_{P_0}$ and $C_{P_0}$ must be considered effective, however,
and deviations of the fitted parameters from the exact asymptotic values are due to
the effective inclusion of neglected higher-order correction terms.

\begin{table}[tb]
  \small
  \caption{Parameters of fits of the form (\ref{eq:polarization_thermal_scaling}) for
    $P_0$ (upper part) resp.\ the form (\ref{thermal_chi_scaling_f}) for $\chi$
    (lower part) to the simulation data. Values in square brackets indicate that the
    corresponding parameter was held fixed in the fit procedure.
    \label{tab:thermal_fits}}
  \begin{tabular}{r@{.}lr@{.}lr@{.}lr@{.}lr@{.}lr@{.}l}\br
    \multicolumn{2}{l}{$A_{P_0}$} & \multicolumn{2}{l}{$B_{P_0}$} &
    \multicolumn{2}{l}{$C_{P_0}$} & \multicolumn{2}{l}{$\beta_c$} &
    \multicolumn{2}{l}{$\rho$} & \multicolumn{2}{l}{$Q$} \\ \mr
    0&8(147)   & $-0$&7(156)  &  $-0$&2(57)  &  0&706(85)   & 0&5(32)   & 0&79 \\
    1&67(35)   & $-1$&59(31)  & [$-0$&5]     &  0&7089(39)  & 0&339(44) & 0&86 \\
    0&736(15)  & $-0$&803(11) & [$-0$&5]     & [0&69315]    & 0&522(33) & 0&14 \\
    0&8088(47) & $-0$&8616(22)& [$-0$&5]     &  0&69499(27) & [0&5]     & 0&23 \\
    0&691(30)  & $-0$&579(78) & $-0$&199(85) &  0&7055(32)  & [0&5]     & 0&85 \\ \mr
   \multicolumn{2}{l}{$A_\chi$} & \multicolumn{2}{l}{$B_\chi$} &
    \multicolumn{2}{l}{$C_\chi$} & \multicolumn{2}{l}{$\beta_c$} &
    \multicolumn{2}{l}{$\rho$} & \multicolumn{2}{l}{$Q$} \\ \mr
       0&5(13)  & 0&15(31)  & [$-1$&0]        &   0&62(13)   & 1&8(18)     & 0&13 \\ 
    $-1$&03(21) & 0&88(12)  & [$-1$&0]        &  [0&69315]   &  0&699(37)  & 0&11 \\
    $-1$&95(11) & 1&549(57) & [$-1$&0]        &   0&7046(19) & [0&5]       & 0&08 \\
    $-2$&2(154) & 2&4(179)  &     0&005(9299) &  [0&69315]   &  0&5(12)    & 0&08 \\
       0&5(24)  & 0&6(12)   &    [0&0]        &   0&647(92)  &  1&2(13)    & 0&13 \\
    $-2$&18(38) & 2&37(27)  &    [0&0]        &  [0&69315]   &  0&520(27)  & 0&12 \\
    $-2$&38(13) & 2&531(66) &    [0&0]        &   0&6944(19) & [0&5]       & 0&12 \\ \br
  \end{tabular}
\end{table}

\subsubsection{Scaling of the polarizability}

\begin{figure}[tb]  
  \centering
  \includegraphics[clip=true,keepaspectratio=true,width=9.5cm]{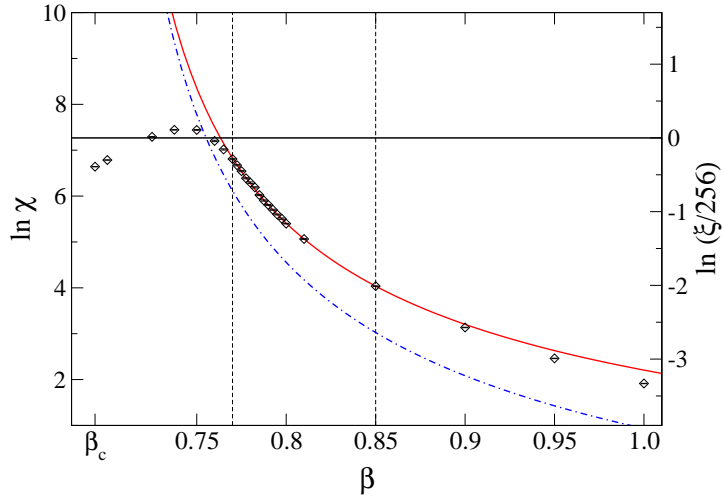}
  \caption
  {Thermal scaling of the polarizability on a $L=256$ lattice. The solid curve shows
    a fit of the functional form (\ref{thermal_chi_scaling_f}) to the data, where the
    parameters $C_\chi=0$ and $\beta_c=\ln 2$ were kept fixed. The vertical dashed
    lines indicate the window of data points included in the fit. To judge the onset
    of finite-size effects, the dashed-dotted curve shows the logarithm of the ratio
    $\xi(\beta,\infty)/L$ from equation (\ref{eq:corrlength}), such that strong size
    effects are expected to appear as soon as $\ln[\xi(\beta,\infty)/L] \gtrsim 0$
    (right scale).}
  \label{reg_F_chi_thermal}
\end{figure}

From the conjecture (\ref{polariz_scaling_conj}) for the near-critical
polarizability, we expect $\chi(\beta)$ to scale analogous to the polarization,
\begin{equation}
  \ln \chi(\beta) = A_\chi+B_\chi(\beta-\beta_c)^{-\rho}+C_\chi\ln(\beta-\beta_c),
  \label{thermal_chi_scaling_f}
\end{equation}
where the differences to the scaling of $P_0$ only show up in the amplitudes
$A_\chi$, $B_\chi$ and $C_\chi=-1$. From the flattening out of the exponential
divergence near $\beta_c$ and by monitoring the quality of fit, we estimate the same
scaling window $\beta_\mathrm{min}=0.77\le\beta\le 0.85=\beta_\mathrm{max}$ for
(\ref{thermal_chi_scaling_f}) we encountered for the polarization. We find fits for
the polarizability to be considerably less stable than those for the polarization,
and we did not succeed to fit all five parameters independently. Fixing $C_\chi=-1$,
a reasonable result for $\rho$ cannot be found, even when additionally fixing
$\beta_c=\ln 2$, cf.\ the data compiled in table \ref{tab:thermal_fits}. Since a fit
with only $\beta_c$ fixed yields $C_\chi\approx 0$, corresponding to an omission of
this correction term, we also tried fits with $C_\chi=0$ fixed, which work
considerably better than fits with $C_\chi=-1$. However, still meaningful results for
$\rho$ and $\beta_c$ can only be found when fixing one of the two parameters, which
then yields good agreement with the asymptotic result. Figure \ref{reg_F_chi_thermal}
shows the simulation data together with a fit with $C_\chi=0$ and $\beta_c=\ln 2$
fixed.  Comparison of the asymptotic correlation length (\ref{eq:corrlength}) with
the system size $L=256$ indicates the approximate onset of finite-size effects as the
critical point is approached.

To see in how far it is possible to distinguish the occurring essential singularity
from a conventional power-law behaviour, we also performed fits to the form
(\ref{thermal_chi_scaling_f}) with the left side replaced by $\chi(\beta)$ instead of
$\ln \chi(\beta)$ and $C_\chi=0$ held fixed. With this power-law form and the same
range of inverse temperatures used for the exponential fits, we arrive at the
following parameters,
\begin{equation}
    \begin{array}{rcl}
    A'_\chi & = & 13.5(49), \\
    B'_\chi & = & 0.040(35), \\
    \beta_c & = & 0.7112(96), \\
    \rho' & = & 3.54(51), \\
    Q & = & 0.13,
  \end{array}
\end{equation}
where $\rho'$ now would correspond to the conventional critical exponent $\gamma$ for
the case of a finite-order phase transition. Thus, in agreement with the experience
from the two-dimensional {\em XY\/} model, power-law fits can be performed with
satisfactory quality if one accepts ``unnaturally'' large exponents such as
$\rho'=3.5$ here.

\section{Conclusions}

We have considered the behaviour of the six-vertex $F$ model on the square-lattice at
its Berezinskii-Kosterlitz-Thouless (BKT) point and within the critical
high-temperature phase with a series of cluster-update Monte Carlo simulations and
subsequent finite-size and thermal scaling analyses. Due to the presence of strong
logarithmic corrections indicated by the exact solution and expected for a theory
with central charge $c=1$, the scaling analysis has to carefully take correction
terms into account and/or treat the presence of (even higher-order) corrections by
omission of simulation points close to the border of the scaling region. Although the
usefulness of the finite-size scaling (FSS) technique has been called into question
at a BKT point due to the occurrence of essential singularities and most studies of
the {\em XY\/} model case solely consider thermal scaling instead \cite{wj:93a}, we
find a FSS analysis for the $F$ model well possible and useful, as long as
corrections to scaling are thoroughly included. The full FSS forms including the
correction terms are explicitly derived from the exact results augmented by the
plausible assumption (\ref{eq:xi_multiplicative_logarithmic}) about the scaling of
the finite-size correlation length. The latter is being confirmed by the analysis of
a combination of observables proportional to a power of the correlation length
without multiplicative logarithmic corrections, providing evidence that the
finite-size correlation length exhibits additive logarithmic corrections in the
present case (as opposed to multiplicative logarithmic corrections such as, e.g., at
the upper critical dimension \cite{kenna:04a}). Due to the ambitious nature of many
of the fits involved, however, one has to cope with the occurrence of competing local
minima of the $\chi^2$ distribution and a distinctive flatness of these minima in
some parameter directions entailed by the slow variation of the logarithmic terms. We
would like to stress that the quality-of-fit parameter $Q$ is found to be not always
sufficient for the detection of neglected higher-order corrections. Omitting the
discussed correction terms, however, the resulting estimates do not even satisfy
moderate expectations of accuracy and are strongly biased. For the FSS analysis the
knowledge of the exact asymptotic critical coupling $\beta_c$ turns out to be highly
beneficial and the results found from the scaling at effective pseudo-critical points
are much less accurate. This might be taken as a caveat for simulations of the {\em
  XY\/} model, where $\beta_c$ is not exactly known.  The correction exponents
$\omega_{P_0}$ for the polarization and $\omega_\chi$ for the polarizability could
not be consistently and accurately determined in fully unrestricted fits, although
constrained fits including further correction terms allow to establish consistency
with the analytical solution.  This experience is shared with simulational studies of
the {\em XY\/} model \cite{wj:97b,hasenbusch:05a}. A thermal scaling analysis of the
low-temperature approach towards criticality does only lead to reasonably precise
results for the present data if at least one of the fit parameters is fixed to its
exact value. A conventional algebraic singularity also fits to the data, but only
when unusually large exponents are accepted.

In addition to the analysis at criticality, we consider the scaling of the
polarization and the polarizability within the critical high-temperature phase. We
find overall good agreement of the outcome with a conjecture
\cite{youngblood,bogoliubov:84a} for the behaviour of the scaling dimension
$x_P(\beta)$ of the polarization in the critical phase, although the resolution of
scaling corrections appears to be even more involved here than at criticality. Close
to the critical point scaling corrections are especially pronounced, since the
scaling dimension $x_P(\beta)$ turns out to have a vertical tangent at $\beta_c$.
This might also contribute to the relatively poor outcome of the FSS analysis of the
peak heights of the polarizability. With respect to the values of the effective
correction exponents $\omega$ found for $\beta<\beta_c$ (cf.\ table
\ref{tab:critical_phase_fits}), we note comparing to the critical point behaviour
that the nature of the corrections seems to be rather different in both cases, such
that the effective correction exponents and amplitudes exhibit fast variation as the
critical phase is entered, which is again related to the singularity of $x_P(\beta)$
at $\beta_c$.

Finally, from deliberately reducing our simulation data set, we note that including
lattice sizes only up to, e.g., $L=128$, most of the estimates for $\beta_c$,
$\gamma/\nu$, $\beta/\nu$, $x_P(\beta)$ and $\rho$ are not found to be compatible
with the asymptotic results in terms of the statistical errors. Thus, consideration
of large system sizes is crucial here for the resolution of scaling corrections, see
also reference~\cite{hasenbusch:05a}. This explains troubles experienced in the
numerical analysis of the $F$ model on a particular, annealed ensemble of fluctuating
quadrangulations, which due to their intrinsic fractality only allow simulations of
lattices with rather small effective linear extents \cite{weigel:04b}. For a more
detailed investigation of the thermal scaling properties an analysis involving
measurements of the finite-size correlation length would be valuable. This, as well
as the examination of the critical phase below the free-fermion point $\beta_f$, is
left to a future investigation.

\ack

This work was partially supported by the EC research network HPRN-CT-1999-00161
``Discrete Random Geometries: from solid state physics to quantum gravity'' and by
the German-Israel-Foundation (GIF) under contract No.\ I-653-181.14/1999. M.W.\
acknowledges support by the DFG through the Graduiertenkolleg ``Quantenfeldtheorie''.
The research at the University of Waterloo was undertaken, in part, thanks to funding
from the Canada Research Chairs Program (Michel Gingras).


\end{document}